\documentclass[11pt]{article}
\usepackage{epsfig}
\def\nn{\nonumber}
\def\be{\begin{equation}}
\def\ee{\end{equation}}
\newcommand{\bea}{\begin{eqnarray}}
\newcommand{\eea}{\end{eqnarray}}
\newcommand{\bdm}{\begin{displaymath}}
\newcommand{\edm}{\end{displaymath}}
\long\def\symbolfootnote[#1]#2{\begingroup%
\def\thefootnote{\fnsymbol{footnote}}\footnote[#1]{#2}\endgroup}

\def\sq2{\sqrt{2}}
\def\drbar{\overline{\rm DR}}

\def\smallP{{\scriptscriptstyle P}}
\def\smallS{{\scriptscriptstyle S}}
\def\smallC{{\scriptscriptstyle C}}

\def\gl{\tilde{g}}
\def\mg{m_{\gl}}
\def\g{\mg^2}
\def\gq{\mg^4}

\def\Veff{V_{\rm eff}}
\def\mix{X}
\def\mixt{\widetilde X}
\def\vs{v_s}
\def\lnb{\overline{\log}\,}

\newcommand{\smallw}{{\scriptscriptstyle W}} %
\def\gbq{\bar g^2}
\def\at{\alpha_t}
\def\ab{\alpha_b}
\def\as{\alpha_s}
\def\oas{{\cal O}(\as)}
\def\oat{{\cal O}(\at)}
\def\oab{{\cal O}(\ab)}
\def\oatas{{\cal O}(\at\as)}
\def\oabas{{\cal O}(\ab\as)}
\def\oatababq{{\cal O}(\at\ab + \ab^2)}
\def\oatqatababq{{\cal O}(\at^2 + \at\ab + \ab^2)}
\def\oatasabas{{\cal O}(\at\as +\ab\as)}
\def\oatq{{\cal O}(\at^2)}

\def\Al{A_\lambda}
\def\Ak{A_\kappa}
\def\As{A_\Sigma}

\def\mt{m_t}

\def\t{\mt^2}
\def\tq{\mt^4}
\def\tu{m_{\tilde{t}_1}^2}
\def\td{m_{\tilde{t}_2}^2}
\def\tul{m_{\tilde{t}_1}}
\def\tdl{m_{\tilde{t}_2}}

\def\sdt{s_{2\theta_t}}
\def\cdt{c_{2\theta_t}}
\def\cdtt{c_{2\bar\theta_t}}
\def\ttbar{\bar \theta_{\tilde t}}

\def\mb{m_b}

\def\cdb{c_{2\theta_b}}

\def\sur{\tilde{u}_R}
\def\sul{\tilde{u}_L}
\def\sdr{\tilde{d}_R}
\def\sdl{\tilde{d}_L}

\def\cptmptt{c_{\varphi_t-\tilde{\varphi}_t}}

\def\DVtu{\frac{\partial \Delta V}{\partial \tu}}
\def\DVtd{\frac{\partial \Delta V}{\partial \td}}
\def\DVcdtq{\frac{\partial \Delta V}{\partial \cdtt^2}}
\def\DVcptmptt{\frac{\partial \Delta V}{\partial \cptmptt}}

\def\DVtt{\frac{\partial^{\,2} \Delta V}{(\partial \mt^2)^2 }}
\def\DVttu{\frac{\partial^{\,2} \Delta V}{\partial \mt^2 \partial \tu }}
\def\DVttd{\frac{\partial^{\,2} \Delta V}{\partial \mt^2 \partial \td }}
\def\DVtutu{\frac{\partial^{\,2} \Delta V}{(\partial \tu)^2 }}
\def\DVtutd{\frac{\partial^{\,2} \Delta V}{\partial \tu \partial \td }}
\def\DVtdtd{\frac{\partial^{\,2} \Delta V}{(\partial \td)^2}}
\def\DVcdtqtu{\frac{\partial^{\,2} \Delta V}{\partial \cdtt^2 \partial \tu }}
\def\DVcdtqtd{\frac{\partial^{\,2} \Delta V}{\partial \cdtt^2 \partial \td }}
\def\DVcdtqt{\frac{\partial^{\,2} \Delta V}{\partial \cdtt^2 \partial \mt^2 }}
\def\DVcdtqcdtq{\frac{\partial^{\,2} \Delta V}{(\partial \cdtt^2)^2 }}

\def\DVtuc2b{\frac{\partial^{\,2} \Delta V}{\partial \tu \partial \cdb^2 }}
\def\DVtdc2b{\frac{\partial^{\,2} \Delta V}{\partial \td \partial \cdb^2 }}
\def\DVtc2b{\frac{\partial^{\,2} \Delta V}{\partial \mt^2 \partial \cdb^2 }}
\def\DVbuc2t{\frac{\partial^{\,2} \Delta V}{\partial \msbu \partial \cdtt^2 }}
\def\DVbdc2t{\frac{\partial^{\,2} \Delta V}{\partial \msbd \partial \cdtt^2 }}
\def\DVbc2t{\frac{\partial^{\,2} \Delta V}{\partial \mb^2 \partial \cdtt^2 }}
\def\DVc2tc2b{\frac{\partial^{\,2} \Delta V}{\partial \cdtt^2 \partial \cdb^2}}

\newenvironment{appendletterA}
 {
  \setcounter{section}{0}
  \setcounter{equation}{0}
  
 }{
 }
\newenvironment{appendletterB}
 {
  \setcounter{equation}{0}
  
 }{
 }
\newenvironment{appendletterC}
 {
  \setcounter{section}{0}
  \setcounter{equation}{0}
  
 }{
 }
\newenvironment{appendletterD}
 {
  \setcounter{equation}{0}
  
 }{
 }

\begin{document}

\begin{titlepage}


{\flushright{
        \begin{minipage}{5cm}
          RM3-TH/09-15 \\
	  LAPTH-1337/09 \\
        \end{minipage}        }

}
\renewcommand{\thefootnote}{\fnsymbol{footnote}}
\vskip 2cm
\begin{center}
\boldmath
{\LARGE\bf On the radiative corrections to the neutral \\[7pt]
Higgs boson masses in the NMSSM}\unboldmath
\vskip 1.cm
{\Large{G.~Degrassi$^{a}$ and P.~Slavich$^{b,\,c}$}}
\vspace*{8mm} \\
{\sl ${}^a$
    Dipartimento di Fisica, Universit\`a di Roma Tre and  INFN, Sezione di
    Roma Tre \\
    Via della Vasca Navale~84, I-00146 Rome, Italy}
\vspace*{2.5mm}\\
{\sl ${}^b$  LAPTH, 9, Chemin de Bellevue, F-74941 Annecy-le-Vieux,  France}
\vspace*{2.5mm}\\
{\sl ${}^c$  LPTHE, 4, Place Jussieu, F-75252 Paris,  France}
\end{center}
\symbolfootnote[0]{{\tt e-mail:}}
\symbolfootnote[0]{{\tt degrassi@fis.uniroma3.it}}
\symbolfootnote[0]{{\tt slavich@lpthe.jussieu.fr}}

\vskip 0.7cm

\begin{abstract}
We provide a full one-loop calculation of the self energies and
tadpoles of the neutral Higgs bosons of the NMSSM. In addition, we
compute the two-loop $\oatasabas$ corrections to the neutral Higgs
boson masses in the effective potential approximation. With respect to
earlier calculations, the newly-computed corrections can account for
shifts of a few GeV in the light scalar and pseudoscalar masses, and
they can also sizeably affect the mixing between singlet and MSSM-like
Higgs scalars. Taking these corrections into account will be crucial
for a meaningful comparison between the MSSM and NMSSM predictions for
the Higgs sector.
\end{abstract}
\vfill
\end{titlepage}    
\setcounter{footnote}{0}


\section{Introduction}

The Next-to-Minimal Supersymmetric extension of the Standard Model, or
NMSSM \cite{nmssm1,nmssm2,nmssmrev}, provides an elegant solution to
the $\mu$ problem of the MSSM \cite{primer}, i.e.~the question of how
to relate the higgsino mass parameter $\mu$ appearing in the
superpotential to the soft SUSY-breaking masses of the other
supersymmetric particles. In the NMSSM the parameter $\mu$ arises as
the vacuum expectation value (vev) of the scalar component of an
additional chiral superfield $S$, singlet with respect to the SM gauge
group and coupled to the MSSM Higgs superfields $H_1$ and $H_2$
through a superpotential term $\lambda\,SH_1H_2\,$. The scalar and
pseudoscalar components of the singlet superfield mix with the MSSM
Higgs fields of matching parity, while the fermion component
(singlino) mixes with the MSSM higgsinos. The quartic Higgs scalar
interaction controlled by the new coupling $\lambda$ can bring the
additional benefit of increasing the tree-level prediction for the
mass of the lightest Higgs boson, which in the MSSM is bounded from
above by the $Z$-boson mass, in contradiction with the lower bound
from direct searches at LEP \cite{LEPhiggs}. This increase allows for
a smaller contribution to the Higgs mass from radiative corrections
involving the top quark and its scalar partner, the stop, thus
reducing the so-called {\em little hierarchy problem} of the MSSM,
i.e.~the need for the stop masses to be substantially larger than the
weak scale. Another interesting feature of the NMSSM is the existence
of scenarios in which a Higgs scalar with SM-like couplings to
fermions and gauge bosons decays mainly into a pair of lighter scalars
or pseudoscalars, requiring a refinement of the strategies for Higgs
searches at the Tevatron and at the LHC \cite{nolose,nonstandard}. A
SM-like Higgs scalar decaying into a pair of pseudoscalars which are
in turn too light to decay into $b$ quarks might even have already
been produced at the LEP, and escaped detection because of its
non-standard decay chain \cite{dermisek}.

Due to the crucial role of radiative corrections in pushing the
prediction for the lightest Higgs boson mass above the LEP bound, an
impressive theoretical effort has been devoted in the past two decades
to the precise determination of the Higgs sector of the
MSSM~\cite{higgsreviews}. After the early
realization~\cite{higgsearly} of the importance of the one-loop $\oat$
corrections\footnote{We define $\alpha_{t,b} = h_{t,b}^2/(4\pi)$,
  where $h_t$ and $h_b$ are the superpotential top and bottom
  couplings, respectively. Here and in the following we denote for
  brevity as $\oat$ the one-loop corrections to the Higgs masses that
  are in fact proportional to $\at m_t^2$, i.e.~$\at^2 v_2^2$. Similar
  abuses of notation affect the other one- and two-loop corrections.}
involving top and stop, full one-loop computations of the MSSM Higgs
masses have been provided \cite{1loop,pbmz}, leading logarithmic
effects at two loops have been included via appropriate
renormalization group equations \cite{rge}, and genuine two-loop
corrections of $\oatas$ \cite{hemphoang,Sven,Zhang,ezhat,dsz}, $\oatq$
\cite{hemphoang,ezhat,bdsz}, $\oabas$ \cite{bdsz2,heidi} and
$\oatababq$ \cite{dds} have been evaluated in the limit of zero
external momentum. All of these corrections have been implemented in
public computer codes \cite{feynhiggs,mssmcodes} for the calculation
of the MSSM mass spectrum. More recently, a nearly complete two-loop
calculation of the MSSM Higgs masses, including electroweak effects
and part of the external momentum dependence, has been
performed~\cite{martin}, and even the leading three-loop effects have
been computed~\cite{3loop}. Finally, a vast literature~\cite{CPV} is
available on the dominant one- and two-loop corrections to the MSSM
Higgs masses in the presence of CP-violating phases in the soft
SUSY-breaking parameters.

In comparison with the case of the MSSM, the computation of the
radiative corrections to the Higgs masses in the NMSSM is not quite as
advanced. The one-loop contributions from diagrams involving top/stop
and bottom/sbottom loops have been computed \cite{nmssm1loop} only in
the effective potential approximation, i.e.~neglecting the external
momentum in the self energies. For what concerns the one-loop
contributions from diagrams involving chargino, neutralino or scalar
loops (the contributions arising from gauge-boson loops are the same
as in the MSSM) only the leading logarithmic terms have been computed
\cite{ewll}. Similarly, among the two-loop contributions only the
leading-logarithmic $\oatas$ and $\oatq$ terms -- borrowed from the
MSSM results under the simplifying assumption of fully degenerate SUSY
masses -- have been taken into account so far.  All of these
corrections have been implemented in a public computer code, {\tt
  NMHDECAY}~\cite{nmhdecay}, which computes masses, couplings and
decay widths of the NMSSM Higgs bosons.

It is clear that, for a proper comparison between the MSSM and NMSSM
predictions and for a precise characterization of the scenarios of
refs.~\cite{nolose,dermisek}, it would be desirable to compute the
masses and mixings in the NMSSM Higgs sector with an accuracy
comparable to that of the calculations implemented in the public MSSM
codes of ref.~\cite{mssmcodes}. In this paper we take a few steps in
this direction. First of all, we provide explicit formulae for the
one-loop corrections to the mass matrices of the neutral CP-even and
CP-odd Higgs bosons. In addition, we compute the two-loop $\oatasabas$
corrections in the approximation of zero external momentum, adapting
to the NMSSM case the techniques (and, in part, the results) developed
for the MSSM in refs.~\cite{dsz,bdsz,dds,ds}. To fully match the
accuracy of the MSSM codes it would also be necessary to include the
two-loop $\oatqatababq$ corrections. These corrections, however,
cannot be straightforwardly adapted from the MSSM case and require a
dedicated calculation. We leave that, as well as a detailed
re-analysis of the NMSSM parameter space taking into account the
improvements in the Higgs mass calculation, to a future publication.

The paper is organized as follows. After this introduction, in section
\ref{sec:general} we describe the Higgs sector of the NMSSM.  In
section \ref{sec:oneloop} we describe the diagrammatic calculation of
the one-loop corrections to the CP-odd and CP-even Higgs mass
matrices.  In section \ref{sec:twoloop} we describe the effective
potential calculation of the two-loop $\oatasabas$ corrections. In
section \ref{sec:results} we show numerical results for some
representative scenarios. In section \ref{sec:concl} we
conclude. Finally, appendix A contains the definitions of the
couplings that enter the calculation of the one-loop radiative
corrections; appendix B provides the explicit expressions for the
one-loop self energies and tadpoles that appear in the computation of
the Higgs boson masses; appendix C contains the formulae for the
two-loop corrections in terms of derivatives of the effective
potential; appendix D contains the definitions of the functions
entering the two-loop results.


\section{The Higgs sector of the NMSSM}
\label{sec:general}
In the SLHA conventions~\cite{slha,slha2} the NMSSM superpotential for
the Higgs superfields $H_1,\,H_2,\,S$ and the quark and lepton
superfields $Q,\, U^c\!\!,\, D^c\!,\,L\,,\,E^c$ reads
\be
\label{superpotential}
W ~=~h_e\, H_1 L E^c + h_d\, H_1 Q D^c + h_u\,Q H_2 U^c  
- \lambda\,S H_1 H_2 + \frac{\kappa}{3}\,S^3~,
\ee
where we neglect colour and generation indices. The $SU(2)$-doublet
superfields are contracted by the antisymmetric tensor
$\epsilon_{ab}$, with $\epsilon_{12} =1$. The corresponding terms in
the soft SUSY-breaking scalar potential are
\bea
\label{soft}
V_{\rm soft} \!\!&=& \!\!m_{H_1}^2H_1^\dagger H_1 + m_{H_2}^2H_2^\dagger H_2 + 
m_{S}^2\,S^* S + m_Q^2 \,Q^\dagger Q + m_L^2 \,L^\dagger L 
+m_U^2 \,\sur^*\sur + m_D^2 \,\sdr^*\sdr + m_E^2 \,\tilde e_R^*\tilde e_R \nn\\
\!\!& +&\!\! \left( h_e\,A_e\,H_1 L \,{\tilde e_R}^*
+ h_d\,A_d\,H_1 Q \,\sdr^* + h_u\,A_u\,Q H_2 \,\sur^* 
-\lambda \,\Al \,S H_1 H_2 
+ \frac{\kappa}{3}\,\Ak \, S^3 ~ + ~{\rm h.c.}\right),
\eea
where for the scalar components of the quark and lepton superfields we
define $\sur = U^{c\,*}$, $\sdr = D^{c\,*}$, $\tilde e_R = E^{c\,*}$,
$Q = (\sul,\sdl)^T$ and $L = (\tilde \nu_L,\tilde e_L)^T$. For
simplicity, we take the soft SUSY-breaking trilinear couplings (as
well as the gaugino masses) to be all real, and we neglect
inter-generational squark and slepton mixing.

The neutral components of the Higgs fields can be decomposed into
their vevs plus their CP-even and CP-odd fluctuations as
\be
\label{fields}
H_i^0 ~=~ v_i + \frac{1}{\sq2}\,(S_i + i P_i)~~(i=1,2)~,~~~~~~~~~
S ~=~ \vs + \frac{1}{\sq2}\,(S_3 + i P_3)~.
\ee
Using the minimization conditions of the tree-level scalar potential
$V_0$ to replace the soft SUSY-breaking Higgs masses
$m_{H_1}^2,\,m_{H_2}^2$ and $m_S^2$ with combinations of the Higgs
vevs and the trilinear couplings, the tree-level mass matrix for the
CP-even fields, $\left({\cal M}^2_S\right)^{\rm tree}$, reads
\be
\label{cpeven}
\left(\begin{array}{ccc}
\gbq v_1^2 + \lambda\,\vs\,\frac{v_2}{v_1}\,\As &
(2\lambda^2-\gbq)v_1 v_2 - \lambda\,\vs\,\As &
2\lambda^2\,v_1\,\vs - \lambda\,v_2\,(\As + \kappa \,\vs)\\
(2\lambda^2-\gbq)v_1 v_2 - \lambda\,\vs\,\As&
\gbq v_2^2 + \lambda\,\vs\,\frac{v_1}{v_2}\,\As&
2\lambda^2\,v_2\,\vs - \lambda\,v_1\,(\As + \kappa \,\vs)\\
2\lambda^2\,v_1\,\vs - \lambda\,v_2\,(\As + \kappa \,\vs)&
2\lambda^2\,v_2\,\vs - \lambda\,v_1\,(\As + \kappa \,\vs)&
\lambda\,\Al\,\frac{v_1 v_2}{\vs}+ \kappa \vs\,(\Ak + 4\,\kappa\,\vs)
\end{array}\right)~,
\ee
where for brevity we define $\As = \Al + \kappa \,\vs$ and $\gbq =
(g^2 + g^{\prime\,2})/2$, $\,g$ and $g^\prime$ being the electroweak
gauge couplings. The CP-even mass matrix is diagonalized by an
orthogonal matrix $R^\smallS$, such that
\be
\label{roteven}
h_i = R^\smallS_{ij}\,S_j~,
\ee
where $h_i$, with $i=1,2,3$, are the CP-even mass eigenstates ordered
by increasing mass.
The tree-level mass matrix for the CP-odd fields, $\left({\cal
  M}^2_P\right)^{\rm tree}$, reads
\be
\label{cpodd}
\left(\begin{array}{ccc}
\lambda\,\vs\,\frac{v_2}{v_1}\,\As &
\lambda\,\vs\,\As &
\lambda\,v_2\,(\As - 3\,\kappa \,\vs)\\
\lambda\,\vs\,\As&
\lambda\,\vs\,\frac{v_1}{v_2}\,\As&
\lambda\,v_1\,(\As -3\, \kappa \,\vs)\\
\lambda\,v_2\,(\As -3\, \kappa \,\vs)&
\lambda\,v_1\,(\As -3\,\kappa \,\vs)&
4\,\lambda\,\kappa\,v_1\,v_2 + \lambda\,\Al\,\frac{v_1 v_2}{\vs} 
-3\,\kappa\,\Ak\,\vs
\end{array}\right)~.
\ee
The CP-odd mass matrix is in turn diagonalized by an orthogonal matrix
$R^\smallP$, such that
\be
\label{rotodd}
a_i = R^\smallP_{ij}\,P_j~,
\ee
where $a_i$ stands for $(G^0,A_1,A_2)$. Here, $G^0$ is the neutral
pseudo-Goldstone boson, while $A_1$ and $A_2$ are the other CP-odd
mass eigenstates, ordered by increasing mass. A commonly adopted
procedure (see, e.g., the first paper in ref.~\cite{nmssm2}) consists
in expressing $P_1$ and $P_2$ in terms of $G^0$ and the orthogonal
combination $A^0$, then defining a new $2\!\times\!2$ orthogonal matrix,
$R^{\smallP\,\prime}$, which rotates $(A^0,P_3)$ into $(A_1,A_2)$. However,
we will follow the SLHA prescription and refrain from this
simplification.

In the neutralino sector, the singlino $\tilde s$ mixes with the
neutral components of the MSSM higgsinos $\tilde h_1^0$ and $\tilde
h_2^0$, which in turn mix with the neutral gauginos $\tilde b$ and
$\tilde w^0$. In the formalism of two-component spinors, the
Lagrangian contains the mass terms
\be
\label{neutralino}
- \frac12\, 
\left(-i \tilde b~-i\tilde w^0~~\tilde h_1^0~~\tilde h_2^0~~\tilde s\right)\,
\left(\begin{array}{ccccc}
M_1 & 0 & -g^\prime\, v_1/\sq2 &   g^\prime\, v_2/\sq2 & 0\\
0 & M_2 & g\, v_1/\sq2 &   -g\, v_2/\sq2 & 0\\
-g^\prime\, v_1/\sq2 &   g^\prime\, v_2/\sq2 & 0 & - \lambda\, \vs &-\lambda\, v_2\\
g \,v_1/\sq2 &   -g \,v_2/\sq2 & - \lambda \,\vs & 0 & -\lambda\, v_1\\
0 & 0 & -\lambda\, v_2 & -\lambda\, v_1 & 2 \,\kappa \,\vs
\end{array}\right)\,\left(\begin{array}{c}
-i \tilde b\\-i\tilde w^0\\\tilde h_1^0\\\tilde h_2^0\\ \tilde s
\end{array}\right)~,
\ee
where $M_1$ and $M_2$ are the soft SUSY-breaking gaugino masses. The 
neutralino mass matrix is diagonalized by a unitary matrix $N$, such that 
\be
\label{diagneut}
\chi^0_i = N_{ij}\, \psi^0_j~,
\ee
where $\psi^0$ stands for $(-i \tilde b,\,-i\tilde w^0,\, \tilde
h_1^0,\,\tilde h_2^0,\,\tilde s)$ and $\chi^0_i$ are the neutralino
mass eigenstates ordered by increasing mass. Under our simplifying
assumptions all the entries of the neutralino mass matrix in
eq.~(\ref{neutralino}) can be taken as real. In this case $N$ becomes
a real matrix, as long as the neutralino mass terms are allowed to
take on negative signs.

Finally, the charged-Higgs and chargino sectors of the NMSSM are not
directly affected by the presence of the singlet superfield. The
expressions for the corresponding mass matrices are the same as in the
MSSM, once we identify
\be
\label{muandtanb}
\tan\beta ~\equiv~ \frac{v_2}{v_1}\,,~~~~~~~
\mu ~\equiv~ \lambda\, \vs\,,~~~~~~~ 
B_\mu ~\equiv~ \lambda\, \vs\,\As - \lambda^2\,v_1 v_2~,
\ee
where $\mu$ is the usual Higgs mass term in the MSSM superpotential
and $B_\mu$ is the corresponding term in the soft SUSY-breaking scalar
potential. In particular, the chargino mass matrix is as in eq.~(22)
of ref.~\cite{slha}, and the charged-Higgs mass is
\be
\label{chargedmass}
m_{H^+}^2 ~=~ \frac{B_\mu}{\sin\beta\cos\beta} ~+~ M_W^2~.
\ee

From eqs.~(\ref{cpeven}), (\ref{cpodd}) and (\ref{neutralino}) it can
be seen that, when $\lambda \ll 1$, the singlet and the
singlino decouple from the Higgs and higgsino sectors of the MSSM,
respectively. Since the experimental lower bounds on the chargino
masses require $\mu$ to be (at least) of the order of 100 GeV,
eq.~(\ref{muandtanb}) implies that, in this limit, $\vs \gg
(v_1,v_2)\,$, so that $B_\mu \approx \lambda\,\kappa\,\vs^2$ is driven
to rather large values -- leading in turn to the so-called decoupling
limit of the MSSM -- unless $\kappa \ll 1$ as well.



\section{One-loop corrections to the Higgs mass matrices}
\label{sec:oneloop}

In this section we describe the calculation of the one-loop
corrections to the neutral Higgs boson masses in the NMSSM. We follow
closely the approach\footnote{There are however a few differences
  between our conventions and those of ref.~\cite{pbmz}: we normalize
  the Higgs vevs in such a way that $(v_1^2+v_2^2) \approx (174$
  GeV$)^2$, and our convention for the sign of the term $\lambda\,S
  H_1H_2$ in the superpotential corresponds to the opposite sign of
  $\mu$ w.r.t.~ref.~\cite{pbmz}.} of the MSSM calculation of
ref.~\cite{pbmz}, which is the one implemented in most public computer
codes \cite{mssmcodes} that compute the mass spectrum of the MSSM.

Including the one-loop corrections in the $\drbar$ renormalization
scheme, and using the minimization conditions of the scalar potential
to replace the soft SUSY-breaking Higgs masses with combinations of
the other parameters, the $3\times3$ mass matrices for the CP-even and
CP-odd fields read
\bea
\label{pieven}
\left({\cal M}^2_S\right)^{\rm 1loop}_{ij}  &=&
\left({\cal M}^2_S\right)^{\rm tree}_{ij} ~+~ 
\frac{1}{\sq2}\,\frac{\delta_{ij}}{v_i}\,T_i 
~-~ \Pi_{s_is_j}(p^2)\\&&\nn\\
\label{piodd}
\left({\cal M}^2_P\right)^{\rm 1loop}_{ij}  &=&
\left({\cal M}^2_P\right)^{\rm tree}_{ij} ~+~
\frac{1}{\sq2}\,\frac{\delta_{ij}}{v_i}\,T_i 
~-~ \Pi_{p_ip_j}(p^2)
\eea
where: the tree-level mass matrices $\left({\cal M}^2_S\right)^{\rm
  tree}$ and $\left({\cal M}^2_P\right)^{\rm tree}\!$ are given in
eqs.~(\ref{cpeven}) and (\ref{cpodd}), respectively, and they are
expressed in terms of $\drbar$-renormalized parameters; $v_i$ stands
for $(v_1,v_2,\vs)$; $T_i$ is the finite part of the one-loop tadpole
diagram for the scalar $S_i$; $\Pi_{s_is_j}(p^2)$ and
$\Pi_{p_ip_j}(p^2)$ are the finite parts of the one-loop self energies
for scalars and pseudoscalars, respectively; $p^2$ is the external
momentum flowing in the self energy. The explicit formulae for the
scalar and pseudoscalar self energies and for the scalar tadpoles are
collected in appendix B. We checked that, in the limit in which
$\lambda \rightarrow 0$ while $\mu\equiv\lambda\,\vs$ is constant, our
results for the $2\!\times\!2$ upper-left submatrix of $\Pi_{s_is_j}$
and for the tadpoles $T_1$ and $T_2$ coincide with the MSSM results of
ref.~\cite{pbmz}, as does the pseudoscalar self energy $\Pi_{AA}$ that
we can obtain by rotating the $2\!\times\!2$ upper-left submatrix of
$\Pi_{p_ip_j}$ by an angle $\beta$.

The radiatively corrected squared mass of the $n$-th scalar or
(physical) pseudoscalar can be obtained by solving iteratively for the
$n$-th eigenvalue of the corresponding mass matrix evaluated at an
external momentum $p^2$ equal to the mass eigenvalue itself (we remark
that this procedure includes in the results for the masses also
contributions that are formally of higher order in the perturbative
expansion). On the other hand, there is a well-known ambiguity in the
definition of the radiatively corrected mixing matrices $R^\smallS$
and $R^\smallP$, because the rotations that diagonalize the
radiatively corrected mass matrices depend on the choice of external
momentum in the self energies. This ambiguity reflects the fact that
the mixing matrices themselves are not physical observables. In our
analysis we will define the radiatively corrected mixing matrix as the
one that diagonalizes the mass matrix at $p^2=0$. This corresponds to
the result obtained in the effective potential approximation.

The tree-level mass matrices in eqs.~(\ref{pieven}) and (\ref{piodd})
depend on the combination of gauge couplings $\gbq$, on the NMSSM
superpotential couplings $\lambda$ and $\kappa$ and on the three vevs
$ v_1,v_2$ and $\vs$. As long as no experimental information on the
parameters of the NMSSM Higgs sector (nor on the validity of the NMSSM
itself) is available, $\lambda,\,\kappa,\,\vs$ and the ratio of vevs
$\tan\beta$ can be considered directly as $\drbar$-renormalized inputs
at some reference scale $Q_0$. On the other hand, the $\drbar$ values of
$v^2 \equiv v_1^2 + v_2^2$, $g$ and $g^{\prime}$ can be extracted from
the experimentally known SM observables. For example, starting from
the muon decay constant $G_\mu$ and the gauge-boson pole masses, we
can make use of the relations
\be
\label{gauge1}
v^{-2} 
~=~ 2\sq2\,G_\mu\,\left(1 -\frac{\Pi^T_{WW}(0)}{M_W^2} - \delta_{\rm VB}\right)~,
\ee
\be
\label{gauge2}
\gbq ~=~ v^{-2}\,M_Z^2\,\left( 1 +
\frac{\Pi^T_{ZZ}(M_Z^2)}{M_Z^2}\right)~,~~~~~~ 
g^2 ~=~ 2\,v^{-2}\,M_W^2\,\left( 1 + \frac{\Pi^T_{WW}(M_W^2)}{M_W^2}\right)~.
\ee
In eqs.~(\ref{gauge1}) and (\ref{gauge2}), $\Pi^T_{VV}(p^2)~~(V =
Z,W)$ denotes the finite and transverse part of the self energy of the
vector bosons, while $\delta_{\rm VB}$ denotes the sum of vertex, box
and wave-function-renormalization corrections to the muon decay
amplitude. The explicit formulae for the vector-boson self energies
are collected in appendix B. The SM contribution to $\delta_{\rm VB}$
was computed long ago \cite{degrassi}, and the SUSY contribution can
be obtained from the MSSM results given in eqs.~(C.13)--(C.22) of
ref.~\cite{pbmz}, by simply extending the sum over the neutralinos to
the five mass eigenstates of the NMSSM.


\section{Two-loop corrections in the effective potential approach}
\label{sec:twoloop}

We now discuss the computation of the two-loop corrections to the
NMSSM Higgs mass matrices in the effective potential approach. The
effective potential for the neutral Higgs sector can be decomposed as
$\Veff = V_0 + \Delta V$, where $\Delta V$ contains the radiative
corrections. The $3\times3$ mass matrices for the CP-even and CP-odd
fields can be decomposed as
\be
\label{massmat}
\left({\cal M}^2_S\right)^{\rm eff}_{ij}  ~=~
\left({\cal M}^2_S\right)^{\rm tree}_{ij} + 
\left(\Delta {\cal M}^2_S\right)_{ij}~,~~~~~~~~~~
\left({\cal M}^2_P\right)^{\rm eff}_{ij} ~=~
\left({\cal M}^2_P\right)^{\rm tree}_{ij} + 
\left(\Delta {\cal M}^2_P\right)_{ij}~,
\ee
and the radiative corrections to the mass matrices are
\bea
\label{massscorr} 
\left(\Delta {\cal M}^2_S\right)_{ij} &=&
-\frac{1}{\sq2}\,\frac{\delta_{ij}}{v_i} 
\left.\frac{\partial \Delta V}{\partial S_i}
\right|_{\rm min} ~+~
\left. \frac{\partial^2 \Delta V}{\partial S_i \partial S_j}
\right|_{\rm min}~,\\
\label{masspcorr} 
\left(\Delta {\cal M}^2_P\right)_{ij} &=&
-\frac{1}{\sq2}\,\frac{\delta_{ij}}{v_i} 
\left.\frac{\partial \Delta V}{\partial S_i}
\right|_{\rm min} ~+~
\left. \frac{\partial^2 \Delta V}{\partial P_i \partial P_j}
\right|_{\rm min}~, 
\eea
where $v_i$ stands for $(v_1,v_2,\vs)$, and the derivatives of the
correction $\Delta V$ are computed at the minimum of $V_{\rm eff}$.
The comparison between eqs.~(\ref{massscorr}) and (\ref{masspcorr})
and eqs.~(\ref{pieven}) and (\ref{piodd}) highlights the
correspondence between tadpoles, self energies and derivatives of the
effective potential.  In the calculation of the MSSM Higgs boson
masses it is customary to reorganize the corrections in such a way
that $\left({\cal M}^2_S\right)^{\rm tree}$ is expressed in terms of
the non-zero eigenvalue of $\left({\cal M}^2_P\right)^{\rm eff}$,
which in the effective potential approximation corresponds to the
physical $A$-boson mass.  In the case of the NMSSM this reorganization
is not as practical, because there are two non-zero eigenvalues of
$\left({\cal M}^2_P\right)^{\rm eff}$. While it is possible to absorb
some of the radiative corrections in an ``effective'' trilinear
coupling $\widetilde \Al$, this parameter does not allow for a direct
physical interpretation. Therefore, we refrain from this manipulation
as well and leave eq.~(\ref{massscorr}) as it stands. Throughout the
calculation we assume that all the parameters entering both the
tree-level and one-loop parts of the mass matrices are renormalized in
the $\drbar$ scheme at a renormalization scale that we denote by $Q$.

The $\oas$ contribution to $\Delta V$ from two-loop diagrams involving
top, stop, gluon and gluino has been computed e.g.~in
refs.~\cite{Zhang,dsz}.  It is the same for the MSSM and for the
NMSSM, and we give it for completeness in appendix C.  The
corresponding $\oatas$ corrections to the mass matrices in
eqs.~(\ref{massscorr}) and (\ref{masspcorr}) can in turn be computed
by exploiting the Higgs-field dependence of the parameters appearing
in $\Delta V$.  As detailed in ref.~\cite{dsz}, if we neglect D-term
contributions controlled by the electroweak gauge couplings the
parameters in the top/top sector depend on the neutral Higgs fields
only through two combinations:
\be
\label{Xes}
X ~\equiv~ |\mix|\,e^{i \varphi} ~=~ h_t \,H_2^0~,~~~~~
\mixt ~\equiv~ |\mixt|\,e^{i \tilde \varphi}~=~
h_t \, \left(A_t\,H_2^0 - \lambda\,S^*\,H_1^{0\,*}\right)\,.
\ee
The top/stop $\oas$ contribution to $\Delta V$ can be expressed in
terms of five field-dependent parameters, which can be chosen as
follows. The squared top and stop masses
\be
\label{topstop}
m_t^2 = |\mix|^2~,~~~~~~~
m^2_{\tilde{t}_{1,2}} = \frac{1}{2} \left[ ( m_Q^2 + m_U^2 + 2\,|\mix|^2\,) \pm 
 \sqrt{ (m_Q^2-m_U^2)^2 + 4 \, |\mixt|^2} \,\right]~, 
\ee
a mixing angle $\ttbar$, with $0\leq \ttbar \leq \pi/2$, which
diagonalizes the stop mass matrix after the stop fields have been
redefined to make it real and symmetric
\be
\label{thetastop}
\sin 2 \,\ttbar = \frac{2\,|\mixt|}{\tu-\td}~,
\ee
and a combination of the phases of $\mix$ and $\mixt$ that we can 
choose as
\be
\label{phases} 
\cos\,(\varphi - \tilde{\varphi}) = 
\frac{ {\rm Re}(\mixt)\,{\rm Re}(\mix) + {\rm Im}(\mixt)\,{\rm Im}(\mix)}
{|\mixt| \, |\mix|} \, .
\ee
A sixth parameter, the gluino mass $\mg$, does not depend on the Higgs
background. In the following we will also refer to $\theta_t$, with
$-\pi/2 <\theta_t<\pi/2$, i.e.~the usual field-independent mixing
angle that diagonalizes the stop mass matrix at the minimum of the
scalar potential.

With a lengthy but straightforward application of the chain rule for
the derivatives of the effective potential, the corrections to the
Higgs mass matrices in eqs.~(\ref{massscorr}) and (\ref{masspcorr})
can be expressed as~\footnote{The differences with respect to
  eqs.~(25)--(30) of ref.~\cite{dsz} have multiple origins: we do not
  absorb part of the corrections in the tree-level mass matrices; we
  adopt the opposite convention for the sign of $\mu$; we take
  directly the derivatives of the renormalized effective potential as
  in refs.~\cite{dds,ds}, removing the need for the
  counterterm-induced shifts $\Delta F_i$ and $\Delta \widetilde
  F_i$.}
\bea
\label{dms11}
\left(\Delta {\cal M}^2_S\right)_{11} & = &
\frac{1}{2} \, h_t^2 \,\mu^2 \,\sdt^2  \,F_3 
~+~ h_t^2\,\tan\beta\,\frac{\mu\,A_t}{\tu-\td}\, F \,,\\
\label{dms12}
\left(\Delta {\cal M}^2_S\right)_{12} & = &
-h_t^2 \,\mu\, m_t\, \sdt \,  F_2 
~-~ \frac{1}{2}\, h_t^2\, A_t \,\mu \, \sdt^2 \, F_3
~-~ h_t^2\,\frac{\mu\,A_t}{\tu-\td}\, F \,, \\
\label{dms22}
\left(\Delta {\cal M}^2_S\right)_{22} & = &
2\, h_t^2\, m_t^2\, F_1
~+~ 2\, h_t^2\, A_t\, m_t\, \sdt\, F_2 
~+~ \frac{1}{2}\, h_t^2\, A_t^2\, \sdt^2\, F_3
~+~ h_t^2\,\cot\beta\,\frac{\mu\,A_t}{\tu-\td}\, F \,,~\\
\label{dms13}
\left(\Delta {\cal M}^2_S\right)_{13} & = &
\frac{1}{2} \, h_t\,\lambda\,m_t \,\mu\,\cot\beta \,\sdt^2  \,F_3 
~-~ h_t\,\lambda\,m_t\,\frac{A_t-2\,\mu\,\cot\beta}{\tu-\td}\, F \,,\\
\label{dms23}
\left(\Delta {\cal M}^2_S\right)_{23} & = &
-h_t\,\lambda \,m_t^2\,\cot\beta \, \sdt \,  F_2 
~-~ \frac{1}{2}\, h_t\,\lambda\, A_t \,m_t\,\cot\beta \, \sdt^2 \, F_3
~-~ h_t\,\lambda\,\cot\beta\frac{m_t\,A_t}{\tu-\td}\, F \,,~~~\\
\label{dms33}
\left(\Delta {\cal M}^2_S\right)_{33} & = &
\frac{1}{2}\, \lambda^2\, m_t^2\,\cot^2\beta\, \sdt^2\, F_3
~+~\lambda^2\,\cot\beta\,\frac{m_t^2\,A_t}{\mu\,(\tu-\td)}\,F\,\\
&&\nn\\
\left(\Delta {\cal M}^2_P\right)_{11} & = & 
h_t^2\,\tan\beta\,\frac{\mu\,A_t}{\tu-\td}\,F_A\,,\\
\left(\Delta {\cal M}^2_P\right)_{12} & = & 
h_t^2\,\frac{\mu\,A_t}{\tu-\td}\,F_A\,,\\
\left(\Delta {\cal M}^2_P\right)_{22} & = & 
h_t^2\,\cot\beta\,\frac{\mu\,A_t}{\tu-\td}\,F_A\,,\\
\left(\Delta {\cal M}^2_P\right)_{13} & = & 
h_t\,\lambda\,\frac{m_t\,A_t}{\tu-\td}\,F_A\,,\\
\left(\Delta {\cal M}^2_P\right)_{23} & = & 
h_t\,\lambda\,\cot\beta\,\frac{m_t\,A_t}{\tu-\td}\,F_A\,,\\
\label{dmp33}
\left(\Delta {\cal M}^2_P\right)_{33} & = & 
\lambda^2\,\cot\beta\,\frac{m^2_t\,A_t}{\mu\,(\tu-\td)}\,F_A\,,
\eea
where the functions $F_i,\,F$ and $F_A$ are combinations of the
derivatives of $\Delta V$ evaluated at the minimum of the effective potential:
\bea
\label{defF1}
F_1 & = &
\DVtt + \DVtutu + \DVtdtd + 2\,\DVttu + 2\,\DVttd + 2\,\DVtutd ~,\\
&&\nn\\
\label{defF2}
F_2 & = &
\DVtutu - \DVtdtd + \DVttu - \DVttd \nn\\
&&
- \frac{4 \,\cdt^2}{\tu-\td}
\left( \DVcdtqt+\DVcdtqtu+\DVcdtqtd\right)~,\\
&&\nn\\
\label{defF3}
F_3 & = &
\DVtutu + \DVtdtd -2\,\DVtutd - \frac{2}{\tu-\td}\left(\DVtu-\DVtd\right)\nn\\
&& + \frac{ 16\, \cdt^2}{(\tu-\td)^2}\,\left( \cdt^2\,\DVcdtqcdtq 
+2\,\DVcdtq \right)
- \frac{ 8\, \cdt^2}{\tu-\td}\,\left( \DVcdtqtu-\DVcdtqtd \right)~,\\
&&\nn\\
F & = & \DVtu - \DVtd - \frac{4\,\cdt^2}{\tu-\td}\,\DVcdtq~,\\
\label{defF}
&&\nn\\
\label{defFA}
F_A & = &  \DVtu - \DVtd - \frac{4\,\cdt^2}{\tu-\td}\,\DVcdtq
- \frac{2\,z_t\,\mu\,\cot\beta}{A_t\,\sdt^2\,(\tu-\td)}\,\DVcptmptt~.
\eea
In eqs.~(\ref{dms11})--(\ref{defFA}) above we adopted the
shortcuts $c_\phi \equiv \cos \phi$ and $s_\phi \equiv \sin \phi$ for
a generic angle $\phi$. The parameters $\mu$ and $\tan\beta$ are
defined in eq.~(\ref{muandtanb}), and $z_t \equiv {\rm
  sign}(A_t-\mu\cot\beta)$. 

At one loop the top and stop contributions to $\Delta V$ depend only
on the corresponding masses. In units of $N_c/(16\pi^2)$, where
$N_c=3$ is a colour factor, the one-loop expressions for the functions
appearing in eqs.~(\ref{dms11})--(\ref{dms33}) are
\be
\label{fi1l1}
F_1^{1\ell} ~=~ \ln\frac{\tu \td}{m_t^4}\,,~~~~
F_2^{1\ell} ~=~ \ln\frac{\tu}{ \td}\,,~~~~
F_3^{1\ell} ~=~ 2 - \frac{\tu+\td}{\tu-\td}\,\ln\frac{\tu}{ \td} \,,
\ee
\be
\label{fi1l2}
F^{1\ell} = F_A^{1\ell} ~=~
\tu\,\left(\log\frac{\tu}{Q^2}-1\right)-
\td\,\left(\log\frac{\td}{Q^2}-1\right)~.
\ee
Inserting eqs.~(\ref{fi1l1}) and (\ref{fi1l2}) in
eqs.~(\ref{dms11})--(\ref{dmp33}) we recover the well-known results
\cite{nmssm1loop} for the one-loop top/stop corrections to the NMSSM
Higgs boson masses in the effective potential approach.

Explicit expressions for the derivatives of the contribution to
$\Delta V$ from two-loop diagrams with top, stop, gluino and gluon are
provided in appendix C.  Rearranging the various terms, it can be
shown that the $2\!\times\!2$ upper-left submatrices of $\Delta {\cal
  M}^2_S$ and $\Delta {\cal M}^2_P$ correspond to the $\oatas$
corrections derived in ref.~\cite{dsz} for the MSSM in the $\drbar$
renormalization scheme. On the other hand, the corrections to the
third row and third column of the mass matrices, which are specific to
the NMSSM, were not previously available. If the one-loop part of the
corrections is expressed in terms of On-Shell (OS) parameters, the
two-loop corrections must be supplemented with counterterm
contributions that account for the shift from $\drbar$ to OS. The
required $\oas$ shifts in the parameters $m_t,\, \tu,\,\td,\,\sdt$ and
$A_t$ can be found in appendix B of ref.~\cite{dsz}.

The computation described above allows us to obtain also the two-loop
$\oabas$ corrections induced by the bottom/sbottom sector, which can
be relevant for large values of $\tan\beta$. To this purpose, the
substitutions $t\rightarrow b$, $\tan\beta \leftrightarrow \cot\beta$,
$\left(\Delta {\cal M}^2_{S,P}\right)_{11} \leftrightarrow ~
\left(\Delta {\cal M}^2_{S,P}\right)_{22}$ and $\left(\Delta {\cal
  M}^2_{S,P}\right)_{13} \leftrightarrow ~ \left(\Delta {\cal
  M}^2_{S,P}\right)_{23}$ must be performed in
eqs.~(\ref{dms11})--(\ref{defFA}). In the case of the bottom/sbottom
corrections, however, passing from the $\drbar$ to the OS scheme
involves additional complications, as explained in ref.~\cite{bdsz2}.

In the case of the MSSM, the computation of the two-loop
$\oatqatababq$ corrections induced by the Yukawa interactions of
quarks, squarks, Higgs bosons and higgsinos is also available
\cite{dds}. In contrast to the case of the $\oatasabas$ corrections,
however, this computation cannot be straightforwardly extended to the
NMSSM, because the Higgs and higgsino sectors are extended by the
presence of the singlet superfield. A dedicated calculation of the
$\oatqatababq$ corrections to the Higgs masses in the NMSSM goes
beyond the scope of this paper.

Finally, since $\Veff$ generates one-particle-irreducible Green's
functions at vanishing external momentum, it is clear that the
effective potential approach neglects the momentum-dependent effects
in the Higgs self energies. The complete computation of the physical
masses of the CP-even and CP-odd Higgs bosons requires the full,
momentum-dependent two-point functions (a detailed discussion of the
correspondence between the effective potential approach and the full
computation has been given in ref.~\cite{bdsz2}). However, in the last
paper of ref.~\cite{martin} it has been shown by direct calculation
that, in the MSSM, the numerical effects of the two-loop
momentum-dependent contributions to the Higgs boson masses are very
small. There is no reason to expect that such effects would be much
larger in the NMSSM.


\section{Numerical examples}
\label{sec:results}
\vspace*{2.5mm}

In this section we briefly discuss the numerical effect of the one-
and two-loop corrections to the NMSSM Higgs masses presented in
sections \ref{sec:oneloop} and \ref{sec:twoloop}, respectively.

Among the Lagrangian parameters that enter the computation of the
NMSSM Higgs masses, the gauge and third-family Yukawa couplings, as
well as the electroweak symmetry breaking parameter $v$, can be
extracted from the known values of various SM observables by taking
into account the appropriate radiative corrections. We use the
following input values for our analysis: the gauge boson masses $M_Z=
91.1876$ GeV and $M_W=80.40$ GeV; the muon decay constant
$G_\mu=1.16637 \times 10^{-5}$ GeV$^{-2}$; the strong coupling
constant $\as(M_Z)=0.1189$; the pole top mass $M_t = 173.1$ GeV; the
running bottom mass $m_b(m_b)=4.23$ GeV; the tau mass $m_\tau=1.777$
GeV.  Consistency with our computation of the one-loop radiative
corrections requires that all the parameters entering the tree-level
mass matrices be expressed in the $\drbar$ renormalization scheme at a
common scale $Q_0$, which we take of the order of the soft
SUSY-breaking scale. For consistency with the computation of the
two-loop $\oatasabas$ corrections, the top and bottom masses and
Yukawa couplings entering the one-loop part of the corrections must
also be expressed in the $\drbar$ scheme.  We determine the running
electroweak gauge couplings and $v$ directly at the scale $Q_0$ by
means of eqs.~(\ref{gauge1}) and (\ref{gauge2}). This procedure
neglects the resummation of potentially large logarithms of the ratio
of the weak scale to the SUSY-breaking scale (incidentally, we also
neglect the small SUSY contributions to $\delta_{\rm VB}$), but it is
accurate enough for the purposes of our study. The top pole mass is
converted into the corresponding running mass, then both the top and
bottom masses are evolved up to the scale $Q_0$ by means of the SM
renormalization group (RG) equations. At that scale the SM running
masses are converted into NMSSM running masses by the inclusion of
gluino-induced threshold corrections (which are the same as in the
MSSM). The tau mass enters only the one-loop part of the calculation
and is not subject to QCD corrections, thus we use directly the pole
mass. Finally, the strong gauge coupling $\as$ enters only the
two-loop part of the calculation, therefore its precise definition
amounts to a higher-order effect. We evolve $\as$ from $M_Z$ to $Q_0$
by means of the SM RG equations.

To exemplify the effect of the one- and two-loop corrections to the
neutral Higgs masses in the NMSSM, we choose the SUSY input parameters
in such a way that the scalar component of the singlet is relatively
light and has a sizeable mixing with the lightest MSSM-like scalar.
For what concerns the Higgs sector, we keep $\lambda$ as a free
parameter and fix the remaining parameters as
\be
\label{parahiggs}
\kappa = \lambda/5~,~~~~
\tan\beta = 2~,~~~~
\Al = 500~{\rm GeV}~,~~~~
\Ak = -10~{\rm GeV}~,~~~~
\mu = 250~{\rm GeV}~,
\ee
where we take $\mu$ as a bookmark for the singlet vev
$\vs=\mu/\lambda$. We adopt a common soft SUSY-breaking mass $M_S$ for
all of the squarks and sleptons, and fix the remaining soft
SUSY-breaking parameters as
\be
\label{parasoft}
A_t=A_b=A_\tau = -1.5\,M_S~,~~~~
M_3= 2\,M_S~,~~~M_2 = 2/3\,M_S~,~~~M_1 = M_S/3~.
\ee
All of the parameters in eqs.~(\ref{parahiggs}) and (\ref{parasoft})
are meant as $\drbar$ running parameters at the scale $Q_0 = M_S$.

\begin{figure}[p]
\begin{center}
\epsfig{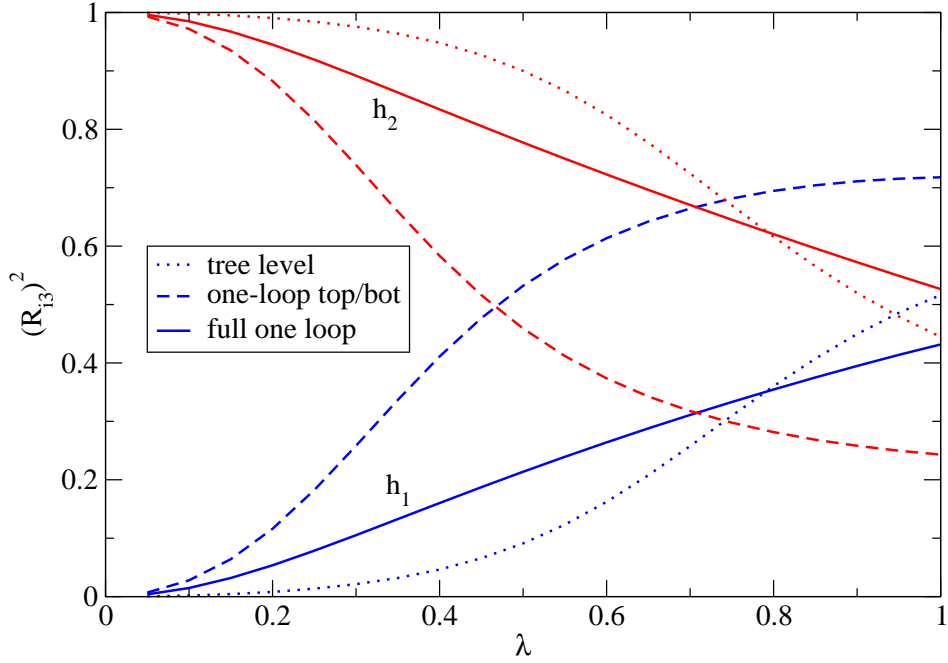}
\end{center}
\vspace*{-4mm}
\caption{The squared rotation matrix element $(R^\smallS_{i3})^2$,
  measuring the singlet component in the scalars $h_1$ and $h_2$, as a
  function of $\lambda$, for $M_S = 300$ GeV. The values of the other
  input parameters and the meaning of the different curves are
  described in the text.}
\label{figrsl}
\end{figure}
\begin{figure}[p]
\begin{center}
\epsfig{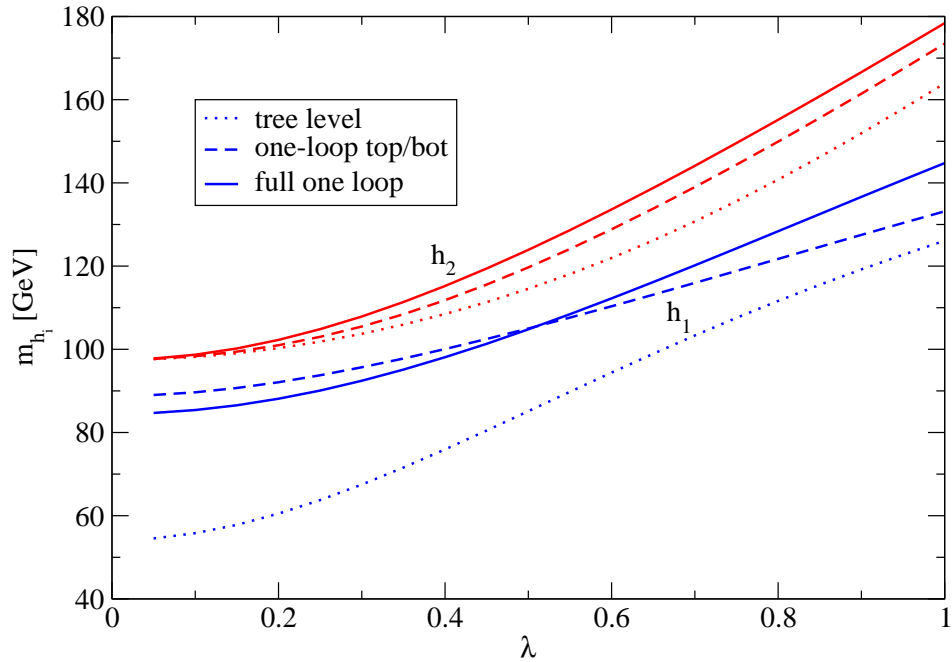}
\end{center}
\vspace*{-4mm}
\caption{The masses of the two lightest scalars $h_1$ and $h_2$ as a
  function of $\lambda$, for $M_S = 300$ GeV. The values of the other
  input parameters and the meaning of the different curves are
  described in the text.}
\label{figmhl}
\end{figure}

\begin{figure}[p]
\begin{center}
\epsfig{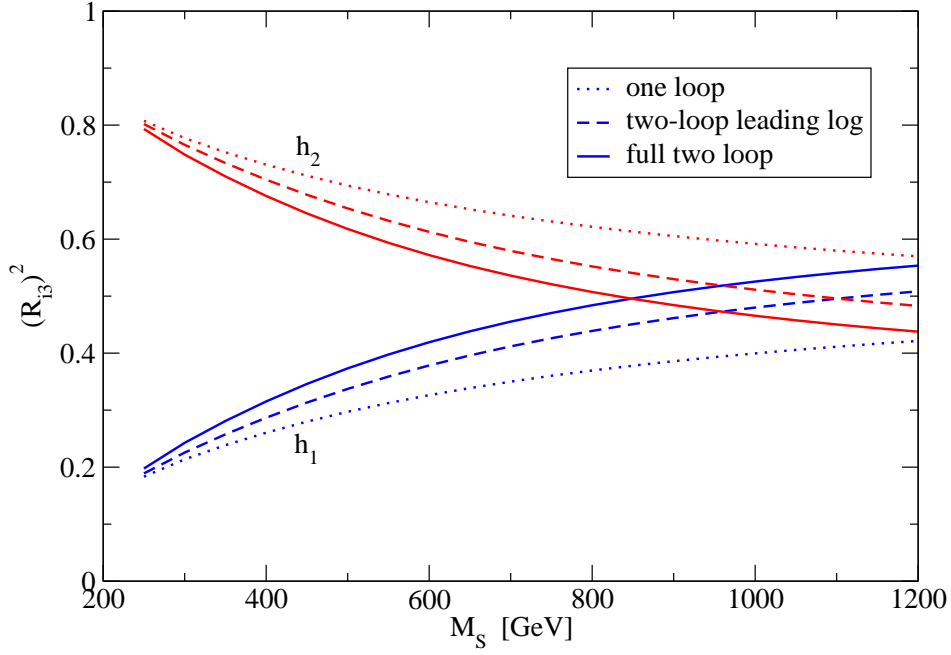}
\end{center}
\vspace*{-4mm}
\caption{The squared rotation matrix element $(R^\smallS_{i3})^2$,
  measuring the singlet component in the scalars $h_1$ and $h_2$, as a
  function of $M_S$, for $\lambda = 0.5$. The values of the other
  input parameters and the meaning of the different curves are
  described in the text.}
\label{figrsl_2}
\end{figure}
\begin{figure}[p]
\begin{center}
\epsfig{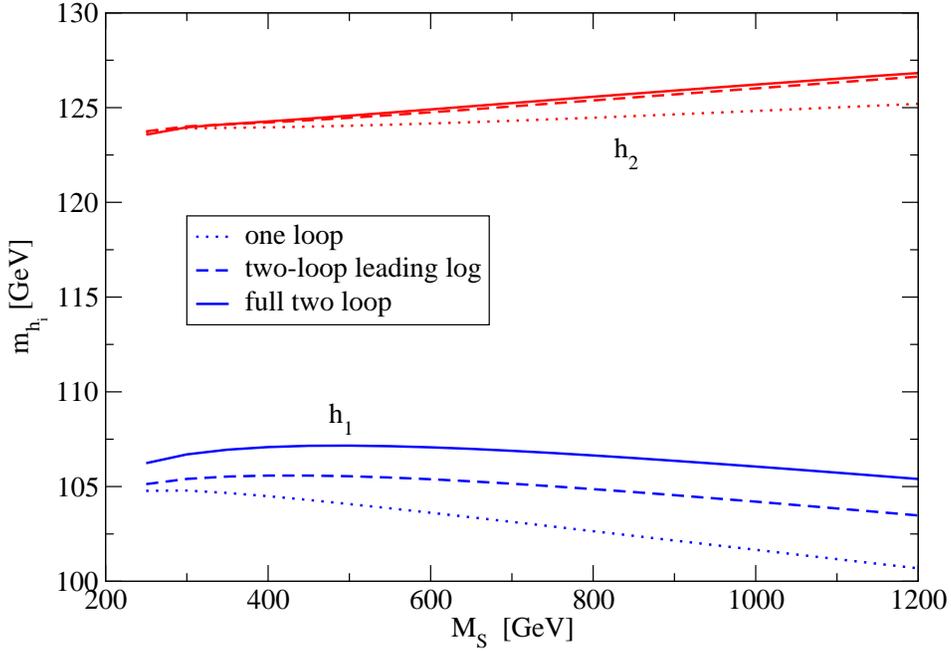}
\end{center}
\vspace*{-4mm}
\caption{The masses of the two lightest scalars $h_1$ and $h_2$ as a
  function of $M_S$, for $\lambda=0.5$. The values of the other
  input parameters and the meaning of the different curves are
  described in the text.}
\label{figmhl_2}
\end{figure}

Figs.~\ref{figrsl} and \ref{figmhl} exemplify the effect of the
one-loop corrections to the NMSSM scalar masses.  In fig.~\ref{figrsl}
we plot the squared rotation matrix elements $(R^\smallS_{13})^2$ and
$(R^\smallS_{23})^2$, which measure the strength of the singlet
component in the two lightest scalars $h_1$ and $h_2$, as a function
of $\lambda$ for $M_S=300$ GeV. In fig.~\ref{figmhl} we plot the
masses of the two lightest scalars for the same choices of inputs.  In
both plots, the dotted lines correspond to the tree-level results; the
dashed lines include the one-loop $\oat$ and $\oab$ corrections
computed in the effective potential approach; finally, the solid lines
correspond to the results of the full one-loop calculation. For the
full one-loop calculation of the rotation matrix the external momentum
in the scalar self energies is set to zero.
It can be seen in fig.~\ref{figrsl} that, at small $\lambda$, the
lightest scalar $h_1$ is dominantly MSSM-like while $h_2$ is
dominantly singlet. When $\lambda$ increases the mixing between
singlet and lightest MSSM-like Higgs increases as well. Meanwhile, the
heaviest scalar $h_3$ has a mass of the order of 600 GeV and its
singlet component is always small. It is interesting to note that --
at least in this point of the parameter space -- the value of
$\lambda$ for which the two lightest mass eigenstates cross over
(i.e., $h_1$ becomes dominantly singlet) depends quite strongly on the
accuracy of the calculation. In particular, when only the quark/squark
contributions to the radiative corrections are included the crossover
occurs for much lower values of $\lambda$ than in the full one-loop
calculation.
Fig.~\ref{figmhl} shows the effect of the radiative corrections to the
two lightest scalar masses. The rise with $\lambda$ in the tree-level
masses is due to the well-known NMSSM contribution to the Higgs
quartic coupling proportional to $\lambda^2\sin^22\beta$. The
comparison between the dotted and dashed lines shows that the $\oat$
corrections induced by top and stop loops have a particularly large
effect on $m_{h_1}$ for small values of $\lambda$, when $h_1$ is light
and mostly MSSM-like. The $\oab$ corrections induced by bottom and
sbottom loops are also included in the dashed lines, but they are
negligible due to the small value of $\tan\beta$.  When $\lambda$
increases the $\oat$ corrections are shared between $m_{h_1}$ and
$m_{h_2}$, and become less relevant due to the increase in the
tree-level masses. However, even for the heavier scalar $h_2$ these
corrections can still amount to several GeV at large
$\lambda$. Finally, the comparison between the solid and dashed lines
in fig.~\ref{figmhl} shows that the remaining one-loop corrections --
which constitute one of the original contributions of this paper --
are also relevant, and can account for shifts of 5--10 GeV in both
masses.

Figs.~\ref{figrsl_2} and \ref{figmhl_2} exemplify the effect of the
two-loop corrections to the NMSSM scalar masses.  In
fig.~\ref{figrsl_2} we plot $(R^\smallS_{13})^2$ and
$(R^\smallS_{23})^2$ as a function of $M_S$ for $\lambda = 0.5$. In
fig.~\ref{figmhl_2} we plot the masses of the two lightest scalars for
the same choices of inputs.  In both plots, the dotted lines
corresponds to the full one-loop results (again, the rotation matrix
is computed at zero external momentum); the dashed lines include the
two-loop leading-logarithmic $\oatas$ contribution to the (2,2) entry
of the scalar mass matrix as implemented in {\tt NMHDECAY}
\cite{nmhdecay}, i.e.
\be
\label{leadlog}
\left(\Delta {\cal M}^2_S\right)_{22}^{\rm LL} ~=~ 
6\,\frac{\at\as}{\pi^2}\,m_t^2\,\log^2\frac{M_S^2}{m_t^2}~.
\ee
Finally, the solid lines correspond to the results of our two-loop
$\oatasabas$ calculation. It can be seen in fig.~\ref{figrsl_2} that
for small $M_S$ the lightest scalar $h_1$ is mostly MSSM-like while
$h_2$ is mostly singlet. When $M_S$ increases, the radiative
corrections increase the mixing between singlet and MSSM-like Higgs.
Fig.~\ref{figmhl_2} shows that the two-loop corrections to the
lightest scalar mass are positive and relatively small. This is a
typical feature of the $\drbar$ computation, in contrast to the OS
computation in which the two-loop corrections are negative and much
larger (for a discussion of this issue in the MSSM see
ref.~\cite{supergroup}). It is interesting to note that, in this
scenario, the leading-logarithmic term accounts only for a fraction
(30\% to 60\%, increasing with $M_S$) of the total $\oatas$
contribution to the (2,2) entry of the scalar mass matrix.  Indeed,
the leading-logarithmic approximation of eq.~(\ref{leadlog}) neglects
potentially large contributions controlled by powers of the ratio
$A_t/M_S$, as well as the possibility of mass splittings among stops
and gluino. The effect of the $\oatas$ corrections to the entries of
the scalar mass matrix other than (2,2) is also non-negligible. The
comparison between the dashed and solid curves for $h_1$ in
figs.~\ref{figrsl_2} and \ref{figmhl_2} shows that, in this point of
the parameter space, the non-leading-logarithmic contributions
contained in our two-loop calculation induce a shift of 1--2 GeV in
$m_{h_1}$, and have a sizeable effect on the mixing matrix as well (on
the other hand, the near overlap of the dashed and solid curves for
$h_2$ in fig.~\ref{figmhl_2} is the result of an accidental
cancellation). One of the attractive features of the NMSSM is the
viability of scenarios in which $M_S$ is not much above the weak
scale.  It is clear from figs.~\ref{figrsl_2} and \ref{figmhl_2} that,
in those scenarios, the leading-logarithmic approximation is not
satisfactory, and a reliable evaluation of the two-loop corrections
requires at least the complete $\oatasabas$ calculation.

\begin{figure}[t]
\begin{center}
\epsfig{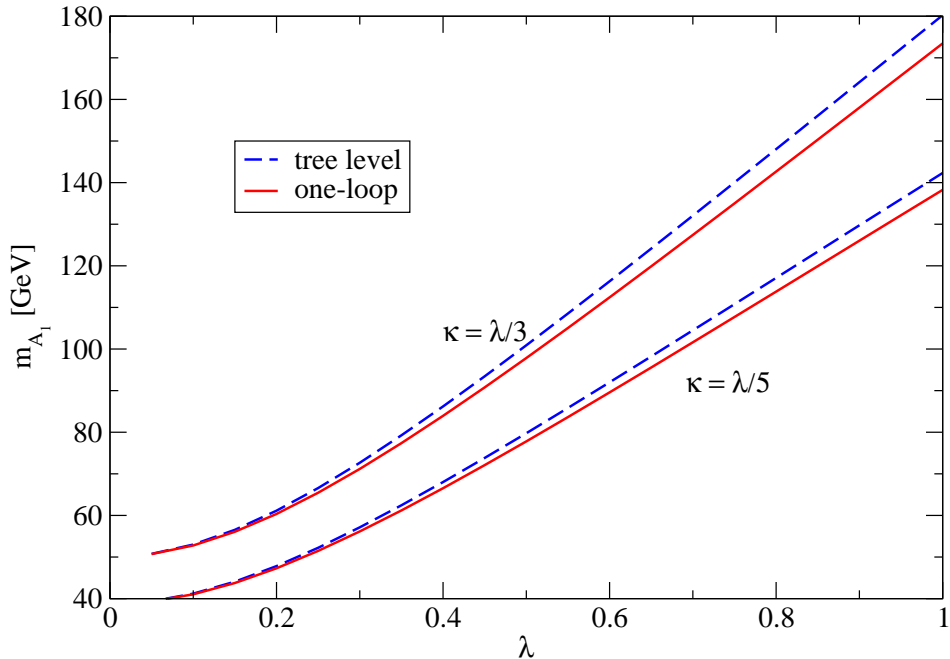}
\end{center}
\vspace*{-5mm}
\caption{The masses of the lightest physical pseudoscalar $A_1$ as a
  function of $\lambda$, for $M_S = 300$ GeV and $\kappa$ set equal to
  either $\lambda/5$ or $\lambda/3$. The values of the other input
  parameters and the meaning of the different curves are described in
  the text.}
\label{figma}
\end{figure}

To conclude this section, we show in fig.~\ref{figma} the effect of
the radiative corrections to the mass of the lightest physical
pseudoscalar $A_1$. For the choice of parameters considered in this
example $A_1$ is almost entirely singlet, therefore its mass is hardly
affected by the one- and two-loop corrections involving quark/squark
loops. The pseudoscalar $A_2$, on the other hand, is almost entirely
MSSM-like, but its tree-level mass is of the order of 600 GeV, thus it
is also not much affected by the radiative corrections. However, the
Higgs self-interactions and the Higgs-higgsino interactions controlled
by the superpotential couplings $\lambda$ and $\kappa$ do induce
non-negligible corrections to the lightest pseudoscalar mass. The
dashed and solid lines in fig.~\ref{figma} correspond to the
tree-level and one-loop determinations of $m_{A_1}$, respectively, as
a function of $\lambda$. The input parameters are chosen as in
eqs.~(\ref{parahiggs}) and (\ref{parasoft}), but we show two sets of
curves corresponding to $\kappa = \lambda/5$ and $\kappa =
\lambda/3$. From the comparison between the dashed and solid curves it
can be seen that the one-loop corrections to $m_{A_1}$ can amount to
several GeV when $\lambda$ and $\kappa$ take on relatively large
values. The two-loop corrections computed in this paper include only
the quark/squark contributions, therefore the corresponding curves
would essentially overlap with the one-loop curves.


\section{Conclusions}
\vspace*{-1mm}
\label{sec:concl}
The NMSSM is an attractive extension of the MSSM: it provides an
elegant solution to the $\mu$ problem, it reduces the need for heavy
superpartners to lift the Higgs mass through radiative corrections and
it has an interesting collider phenomenology. However, the accuracy of
the theoretical predictions for the NMSSM Higgs masses has until now
been stuck to the level that for the MSSM had been achieved in the
mid-1990s. In this paper we took a few steps towards bridging the
accuracy gap between the NMSSM and MSSM calculations. In particular,
we provided a full one-loop calculation of the self energies and
tadpoles of the neutral Higgs bosons of the NMSSM, and we computed the
two-loop $\oatasabas$ corrections to the neutral Higgs boson masses in
the effective potential approximation. We showed that both classes of
corrections can induce shifts of a few GeV in the light scalar and
pseudoscalar masses, and they can also sizeably affect the mixing
between singlet and MSSM-like Higgs scalars. Taking these corrections
into account in phenomenological analyses of the NMSSM Higgs sector
will be crucial for a meaningful comparison between the MSSM and NMSSM
predictions.

\section*{Acknowledgments}

We thank F.~Boudjema for discussions. This work was supported in part
by an EU Marie-Curie Research Training Network under contract
MRTN-CT-2006-035505 and by ANR under contract BLAN07-2\_194882.



\section*{Appendix A: definitions of the couplings}
\begin{appendletterA}

In this appendix we provide explicit formulae for the couplings that
enter the calculation of the one-loop corrections to the Higgs boson
masses in the NMSSM.

\paragraph{Higgs--sfermion couplings:}

the terms in the NMSSM Lagrangian relevant to the calculation of the
sfermion contributions to the Higgs self energies can be written as
\be
\label{higgssfe}
{\cal L} ~\supset~ 
-\sum_{ijk\ell} \lambda_{s_is_j\tilde F_k\tilde F_\ell} 
~S_i S_j \tilde F^*_k \tilde F_\ell
~-\sum_{ijk\ell} \lambda_{p_ip_j\tilde F_k\tilde F_\ell} 
~P_i P_j \tilde F^*_k \tilde F_\ell
~- \sum_{ik\ell} \lambda_{s_i\tilde F_k\tilde F_\ell}
~S_i \tilde F^*_k \tilde F_\ell
~- \sum_{ik\ell} i\,\lambda_{p_i\tilde F_k\tilde F_\ell}
~P_i \tilde F^*_k \tilde F_\ell~,
\ee
where $\tilde F_i = (\tilde f_L,\tilde f_R)$ represent the sfermions
in the basis of interaction eigenstates.  The quartic couplings in
eq.~(\ref{higgssfe}) are symmetric with respect to the exchange of $i$
and $j$ as well as with respect to the exchange of $k$ and $\ell$. The
trilinear couplings of the scalars are symmetric with respect to the
exchange of $k$ and $\ell$, whereas the trilinear couplings of the
pseudoscalars are antisymmetric. The couplings to up-type and
down-type squarks (the generalization to the sleptons is
straightforward) are
\[
\lambda_{s_1s_1\tilde U_1\tilde U_1} ~=~
\lambda_{p_1p_1\tilde U_1\tilde U_1} ~=~
 \frac{\gbq}{2}\,g_{u_L}\,,~~~~~~
\lambda_{s_1s_1\tilde U_2\tilde U_2} ~=~
\lambda_{p_1p_1\tilde U_2\tilde U_2} ~=~
\frac{\gbq}{2}\,g_{u_R}\,,
\]
\[
\lambda_{s_2s_2\tilde U_1\tilde U_1} ~=~ 
\lambda_{p_2p_2\tilde U_1\tilde U_1} ~=~ 
-\frac{\gbq}{2}\,g_{u_L} ~+~ \frac{h_u^2}{2}\,,~~~~~~
\lambda_{s_2s_2\tilde U_2\tilde U_2} ~=~ 
\lambda_{p_2p_2\tilde U_2\tilde U_2} ~=~ 
-\frac{\gbq}{2}\,g_{u_R} ~+~ \frac{h_u^2}{2}\,,
\]
\be
\label{hsfquau}
\lambda_{s_1s_3\tilde U_1\tilde U_2} ~=~ 
-\lambda_{p_1p_3\tilde U_1\tilde U_2} ~=~ -\frac{h_u\,\lambda}{4}~,
\ee
\[
\lambda_{s_1s_1\tilde D_1\tilde D_1} ~=~ 
\lambda_{p_1p_1\tilde D_1\tilde D_1} ~=~ 
\frac{\gbq}{2}\,g_{d_L} ~+~ \frac{h_d^2}{2}\,,~~~~~~
\lambda_{s_1s_1\tilde D_2\tilde D_2} ~=~ 
\lambda_{p_1p_1\tilde D_2\tilde D_2} ~=~ 
\frac{\gbq}{2}\,g_{d_R} ~+~ \frac{h_d^2}{2}\,,
\]
\[
\lambda_{s_2s_2\tilde D_1\tilde D_1} ~=~ 
\lambda_{p_2p_2\tilde D_1\tilde D_1} ~=~ 
-\frac{\gbq}{2}\,g_{d_L}\,,~~~~~~
\lambda_{s_2s_2\tilde D_2\tilde D_2} ~=~ 
\lambda_{p_2p_2\tilde D_2\tilde D_2} ~=~ 
-\frac{\gbq}{2}\,g_{d_R}\,,
\]
\be
\label{hsfquad}
\lambda_{s_2s_3\tilde D_1\tilde D_2} ~=~
-\lambda_{p_2p_3\tilde D_1\tilde D_2} ~=~ -\frac{h_d\,\lambda}{4}~,
\ee
\[
\lambda_{s_1\tilde U_1\tilde U_1} ~=~ \sq2\,\gbq\,g_{u_L}\,v_1\,,~~~~~~
\lambda_{s_1\tilde U_2\tilde U_2} ~=~ \sq2\,\gbq\,g_{u_R}\,v_1\,,
\]
\[
\lambda_{s_2\tilde U_1\tilde U_1} 
~=~ -\sq2\,\gbq\,g_{u_L}\,v_2 ~+~ \sq2\,h_u^2\,v_2\,,~~~~~~
\lambda_{s_2\tilde U_2\tilde U_2} 
~=~ -\sq2\,\gbq\,g_{u_R}\,v_2 ~+~ \sq2\,h_u^2\,v_2\,,
\]
\be
\label{hsftriu}
\lambda_{s_1\tilde U_1\tilde U_2} =
\lambda_{p_1\tilde U_2\tilde U_1} \,=\, -\frac{h_u\,\vs\,\lambda}{\sq2}\,,~~~~
\lambda_{s_2\tilde U_1\tilde U_2} =
\lambda_{p_2\tilde U_1\tilde U_2} \,=\, \frac{h_u\,A_u}{\sq2}\,,~~~~
\lambda_{s_3\tilde U_1\tilde U_2} =
\lambda_{p_3\tilde U_2\tilde U_1} \,=\, -\frac{h_u\,v_1\,\lambda}{\sq2}~,
\ee
\[
\lambda_{s_1\tilde D_1\tilde D_1} ~=~ 
\sq2\,\gbq\,g_{d_L}\,v_1 ~+~ \sq2\,h_d^2\,v_1\,,~~~~~~
\lambda_{s_1\tilde D_2\tilde D_2} 
~=~ \sq2\,\gbq\,g_{d_R}\,v_1 ~+~ \sq2\,h_d^2\,v_1\,,
\]
\[
\lambda_{s_2\tilde D_1\tilde D_1} 
~=~ -\sq2\,\gbq\,g_{d_L}\,v_2\,,~~~~~~
\lambda_{s_2\tilde D_2\tilde D_2} 
~=~ -\sq2\,\gbq\,g_{d_R}\,v_2\,,
\]
\be
\label{hsftrid}
\lambda_{s_1\tilde D_1\tilde D_2} =
\lambda_{p_1\tilde D_1\tilde D_2} \,=\, \frac{h_d\,A_d}{\sq2}\,,~~~~
\lambda_{s_2\tilde D_1\tilde D_2} =
\lambda_{p_2\tilde D_2\tilde D_1} \,=\, -\frac{h_d\,\vs\,\lambda}{\sq2}\,,~~~~
\lambda_{s_3\tilde D_1\tilde D_2} =
\lambda_{p_3\tilde D_2\tilde D_1} \,=\, -\frac{h_d\,v_2\,\lambda}{\sq2}~,
\ee
where $g_f = I_3^f - e_f\,\sin^2\theta_W$. Here $I_3^f$ is the weak
isospin and $e_f$ is the electric charge of the chiral superfield that
contains the sfermion (e.g.,~$e_{u_R} = -2/3$). The couplings that
cannot be obtained by swapping the last two indices and -- for the
quartic couplings -- the first two indices of those in
eqs.~(\ref{hsfquau})--(\ref{hsftrid}) are all vanishing.

In the absence of CP-violating phases in the sfermion mass matrix, the
sfermion mass eigestates $\tilde f_i = (\tilde f_1,\tilde f_2)$ are
related to the interaction eigenstates $\tilde F_i$ by an orthogonal
rotation:
\be \tilde f_i = R^f_{ij} \,\tilde F_j~,~~~~~~~~ R^f
~=~\left(\begin{array}{rc}\cos\theta_f&\sin\theta_f\\ -\sin\theta_f&\cos\theta_f\end{array}\right)~,
  \ee
where $\theta_f$ is the mixing angle of the sfermions $\tilde f$. The
quartic and trilinear couplings between Higgs fields and sfermion mass
eigenstates are related to the corresponding couplings between Higgs
fields and sfermion interaction eigenstates as follows
\be
\label{rotsf}
\lambda_{\phi_i\phi_j\tilde f_k\tilde f_\ell} = 
R^f_{ka}\,R^f_{\ell b}
\,\lambda_{\phi_i\phi_j\tilde F_a\tilde F_b}~,~~~~~~~
\lambda_{\phi_i\tilde f_k\tilde f_\ell} = 
R^f_{ka}\,R^f_{\ell b}
\,\lambda_{\phi_i\tilde F_a\tilde F_b}~.
\ee
where $\phi_i$ represents either $s_i$ or $p_i$, and summation over repeated
indices is understood.

\paragraph{Higgs self-couplings:}
the terms in the NMSSM Lagrangian relevant to the calculation of the
neutral-Higgs contributions to the Higgs self energies can be written as
\bea
{\cal L} &\supset& 
-\sum_{ijk\ell} \lambda_{s_is_j s_k s_\ell} 
~S_i S_j  S_k  S_\ell
~-\sum_{ijk\ell} \lambda_{p_ip_j p_k p_\ell} 
~P_i P_j  P_k  P_\ell 
~-\sum_{ijk\ell} \lambda_{s_is_j p_k p_\ell} 
~S_i S_j  P_k  P_\ell\nn\\
\label{selfhiggs}
&& - \sum_{ik\ell} \lambda_{s_i s_k s_\ell}
~S_i  S_k  S_\ell
~- \sum_{ik\ell} \lambda_{s_i p_k p_\ell}
~S_i  P_k  P_\ell~.
\eea
The quartic and trilinear neutral-Higgs self couplings entering
eq.~(\ref{selfhiggs}) are symmetric with respect to the permutation of
any two indices corresponding to fields of the same parity. They read
\[
\lambda_{s_1s_1s_1s_1} ~=~ \lambda_{s_2s_2s_2s_2} ~=~ 
\lambda_{p_1p_1p_1p_1} ~=~ \lambda_{p_2p_2p_2p_2} ~=~\frac{\gbq}{16}~,~~~~~~
\lambda_{s_3s_3s_3s_3} ~=~ \lambda_{p_3p_3p_3p_3} ~=~\frac{\kappa^2}{4}~,
\]
\[
\lambda_{s_1s_1s_2s_2} ~=~ \lambda_{p_1p_1p_2p_2} ~=~ 
\frac{1}{48}\,\left(2\lambda^2-\gbq\right)~,~~~~~~
\lambda_{s_1s_2s_3s_3} ~=~ \lambda_{p_1p_2p_3p_3} ~=~-\frac{\lambda\kappa}{24}~,
\]
\[
\lambda_{s_1s_1s_3s_3} ~=~ \lambda_{s_2s_2s_3s_3} ~=~ 
\lambda_{p_1p_1p_3p_3} ~=~ \lambda_{p_2p_2p_3p_3} ~=~ 
\frac{\lambda^2}{24}~,
\]
\[
\lambda_{s_1s_1p_1p_1} ~=~ \lambda_{s_2s_2p_2p_2} ~=~ \frac{\gbq}{8}~,~~~~~~
\lambda_{s_1s_1p_2p_2} ~=~ \lambda_{s_2s_2p_1p_1} ~=~
\frac{1}{8}\,\left(2\lambda^2-\gbq\right)~,
\]
\[
\lambda_{s_1s_1p_3p_3} ~=~ \lambda_{s_2s_2p_3p_3} ~=~ 
\lambda_{s_3s_3p_1p_1} ~=~ \lambda_{s_3s_3p_2p_2} ~=~ 
\frac{\lambda^2}{4}~,~~~~~~
\]
\be
\label{selfquartic}
\lambda_{s_1s_2p_3p_3} ~=~ \lambda_{s_3s_3p_1p_2} ~=~ 
-\lambda_{s_1s_3p_2p_3} ~=~ -\lambda_{s_2s_3p_1p_3} ~=~ 
\frac{\lambda\,\kappa}{4}~,~~~~~~
\lambda_{s_3s_3p_3p_3} ~=~ \frac{\kappa^2}{2}~,
\ee

\[
\lambda_{s_1s_1s_1} ~=~ \lambda_{s_1p_1p_1} ~=~ \frac{\gbq \,v_1}{2\sq2}~,~~~~~~
\lambda_{s_2s_2s_2} ~=~ \lambda_{s_2p_2p_2} ~=~ \frac{\gbq \,v_2}{2\sq2}~,
\]
\[
\lambda_{s_1p_2p_2} ~=~ 3\,\lambda_{s_1s_2s_2} ~=~  
\frac{v_1}{2\sq2}\,\left(2\lambda^2-\gbq\right)~,~~~~~~
\lambda_{s_2p_1p_1} ~=~ 3\,\lambda_{s_2s_1s_1} ~=~  
\frac{v_2}{2\sq2}\,\left(2\lambda^2-\gbq\right)~,
\]
\[
\lambda_{s_3p_1p_1} ~=~ \lambda_{s_3p_2p_2} ~=~
3\,\lambda_{s_3s_1s_1} ~=~ 3\,\lambda_{s_3s_2s_2} ~=~ \frac{\lambda^2 \,\vs}{\sq2}~,
\]
\[
\lambda_{s_3s_3s_3} ~=~ \frac{\kappa \,\Ak}{3\sq2} + \sq2\,\kappa^2\,\vs ~,~~~~~~
\lambda_{s_3p_3p_3} ~=~ -\frac{\kappa \,\Ak}{\sq2} + \sq2\,\kappa^2\,\vs ~,
\]
\[
\lambda_{s_1s_3s_3} ~=~
\frac{\lambda}{3\sq2}\,\left(\lambda \,v_1-\kappa v_2\right)~,~~~~~~
\lambda_{s_2s_3s_3} ~=~
\frac{\lambda}{3\sq2}\,\left(\lambda \,v_2-\kappa v_1\right)~,~~~~~~
\]
\[
\lambda_{s_1p_3p_3} ~=~
\frac{\lambda}{\sq2}\,\left(\lambda \,v_1+\kappa v_2\right)~,~~
\lambda_{s_2p_3p_3} ~=~
\frac{\lambda}{\sq2}\,\left(\lambda \,v_2+\kappa v_1\right)~,~~~
\lambda_{s_3p_1p_3} ~=~ -\frac{\lambda\,\kappa\,v_2}{\sq2}~,~~
\lambda_{s_3p_2p_3} ~=~ -\frac{\lambda\,\kappa\,v_1}{\sq2}~,
\]
\be
\label{selftrilinear}
\lambda_{s_1s_2s_3} ~=~
-\frac{\lambda \,\Al}{6\sq2}- \frac{\lambda\,\kappa\,\vs}{3\sq2} ~,~~~~~~
\lambda_{s_1p_2p_3} ~=~ \lambda_{s_2p_1p_3} ~=~ 
\frac{\lambda \,\Al}{2\sq2}- \frac{\lambda\,\kappa\,\vs}{\sq2} ~,~~~~~~
 \lambda_{s_3p_1p_2} ~=~
\frac{\lambda \,\Al}{2\sq2}+ \frac{\lambda\,\kappa\,\vs}{\sq2} ~.
\ee
All of the couplings that cannot be obtained by permuting the indices
of the couplings in eqs.~(\ref{selfquartic}) and (\ref{selftrilinear})
vanish.
Rotating two scalar or pseudoscalar interaction eigenstates into mass
eigenstates as in eqs.~(\ref{roteven}) and (\ref{rotodd}), and
exploiting the permutation symmetry of the original couplings, the
couplings that enter the calculation of the scalar self energies can
be expressed as
\[
\lambda_{s_is_jh_kh_\ell} ~=~ 6\,
R^\smallS_{ka}\,R^\smallS_{\ell b}
\,\lambda_{s_is_js_as_b}~,~~~~~
\lambda_{s_is_ja_ka_\ell} ~=~ 
R^\smallP_{ka}\,R^\smallP_{\ell b}
\,\lambda_{s_is_jp_ap_b}~,
\]
\be
\label{selfscalrot}
\lambda_{s_ih_kh_\ell} ~=~  3\,
R^\smallS_{ka}\,R^\smallS_{\ell b}
\,\lambda_{s_is_as_b}~,~~~~~
\lambda_{s_ia_ka_\ell} ~=~ 
R^\smallP_{ka}\,R^\smallP_{\ell b}
\,\lambda_{s_ip_ap_b}~.
\ee
Similarly, the couplings that enter the calculation of
the pseudoscalar self energies are
\[
\lambda_{p_ip_ja_ka_\ell} ~=~ 6\,
R^\smallP_{ka}\,R^\smallP_{\ell b}
\,\lambda_{p_ip_jp_ap_b}~,~~~~~
\lambda_{p_ip_jh_kh_\ell} = 
R^\smallS_{ka}\,R^\smallS_{\ell b}
\,\lambda_{s_as_bp_ip_j}~,
\]
\be
\label{selfpseurot}
\lambda_{p_ia_kh_\ell} ~=~
2\,R^\smallS_{\ell a}\,R^\smallP_{k b}
\,\lambda_{s_a p_bp_i}~.
\ee

The terms in the NMSSM Lagrangian relevant to the calculation of the
charged-Higgs contributions to the Higgs self energies can be written
as
\bea
{\cal L} &\supset&
-\sum_{ijk\ell} \lambda_{s_is_j h^+_k h^-_\ell} 
~S_i S_j  h^+_k  h^-_\ell
~-\sum_{ijk\ell} \lambda_{p_ip_j h^+_k h^-_\ell} 
~P_i P_j  h^+_k  h^-_\ell\nn\\
&&- \sum_{ik\ell} \lambda_{s_i h^+_k h^-_\ell}
~S_i  h^+_k  h^-_\ell
~- \sum_{ik\ell} i\,\lambda_{p_i h^+_k h^-_\ell}
~P_i  h^+_k  h^-_\ell~,
\label{charhiggs}
\eea
where we express the charged Higgs fields directly in the basis of
mass eigenstates, so $h_i^\pm$ stands for $(G^\pm,H^\pm)$. In our
conventions the relation between the charged Higgs mass eigenstates
and interaction eigenstates reads~\footnote{This differs from the
  conventions of Ref.~\cite{pbmz} by one field redefinition.}
\be
\label{rothpm}
\left(\begin{array}{c}G^\pm\\H^\pm\end{array}\right)
~=~ R^\smallC\,
\left(\begin{array}{c}H_1^\pm\\H_2^\pm\end{array}\right)~,~~~~~
R^\smallC ~=~ \left(\begin{array}{rc}
-\cos\beta&\sin\beta\\\sin\beta&\cos\beta
\end{array}\right)~.
\ee
The quartic couplings in eq.~(\ref{charhiggs}) are symmetric with
respect to the exchange of $i$ and $j$ as well as with respect to the
exchange of $k$ and $\ell$. The trilinear couplings of the scalars are
symmetric with respect to the exchange of $k$ and $\ell$, whereas the
trilinear couplings of the pseudoscalars are antisymmetric. Explicit
expressions for the couplings are
\[
\lambda_{s_1s_1h_1^+h_1^-} ~=~ \lambda_{s_2s_2h_2^+h_2^-} ~=~ 
\lambda_{p_1p_1h_1^+h_1^-} ~=~ \lambda_{p_2p_2h_2^+h_2^-} ~=~
\frac18\,(g^2 + g^{\prime\,2}\cos 2\beta)~,
\]
\[
\lambda_{s_1s_1h_2^+h_2^-} ~=~ \lambda_{s_2s_2h_1^+h_1^-} ~=~ 
\lambda_{p_1p_1h_2^+h_2^-} ~=~ \lambda_{p_2p_2h_1^+h_1^-} ~=~
\frac18\,(g^2 - g^{\prime\,2}\cos 2\beta)~,
\]
\[
\lambda_{s_1s_2h_1^+h_1^-} ~=~ \lambda_{p_1p_2h_2^+h_2^-} ~=~ 
-\lambda_{s_1s_2h_2^+h_2^-} ~=~ -\lambda_{p_1p_2h_1^+h_1^-} ~=~
\frac18\,(2\lambda^2 - g^2)\,\sin 2\beta~,
\]
\[
\lambda_{s_1s_1h_1^+h_2^-} ~=~ \lambda_{p_1p_1h_1^+h_2^-} ~=~ 
-\lambda_{s_2s_2h_1^+h_2^-} ~=~ -\lambda_{p_2p_2h_1^+h_2^-} ~=~
-\frac{\;g^{\prime\,2}}{8}\,\sin 2\beta~,
\]
\[
\lambda_{s_1s_2h_1^+h_2^-} ~=~ -\lambda_{p_1p_2h_1^+h_2^-} ~=~ 
\frac18\,(2\lambda^2 - g^2)\,\cos 2\beta~,~~~~~
\lambda_{s_3s_3h_1^+h_2^-} ~=~ -\lambda_{p_3p_3h_1^+h_2^-} ~=~ 
-\frac{\kappa\lambda}{2}\,\cos 2\beta~,
\]
\be
\label{chaquartic}
\lambda_{s_3s_3h_1^+h_1^-} ~=~ \lambda_{p_3p_3h_2^+h_2^-} ~=~
\frac{\lambda}{2}\,(\lambda - \kappa\,\sin 2\beta)~,~~~~~
\lambda_{s_3s_3h_2^+h_2^-} ~=~ \lambda_{p_3p_3h_1^+h_1^-} ~=~
\frac{\lambda}{2}\,(\lambda + \kappa\,\sin 2\beta)~,
\ee

\[
\lambda_{s_1h_1^+h_1^-} ~=~\frac{1}{2\sq2}\,\left(
v_1\,(g^2+g^{\prime\,2}\cos2\beta)~+~v_2\,(2\lambda^2-g^2)\,\sin2\beta\right)~,
\]
\[
\lambda_{s_1h_2^+h_2^-} ~=~\frac{1}{2\sq2}\,\left(
v_1\,(g^2-g^{\prime\,2}\cos2\beta)~-~v_2\,(2\lambda^2-g^2)\,\sin2\beta\right)~,
\]
\[
\lambda_{s_1h_1^+h_2^-} ~=~\frac{1}{2\sq2}\,\left(
-v_1\,g^{\prime\,2}\sin2\beta~+~v_2\,(2\lambda^2-g^2)\,\cos2\beta\right)~,
\]
\[
\lambda_{s_2h_1^+h_1^-} ~=~\frac{1}{2\sq2}\,\left(
v_2\,(g^2-g^{\prime\,2}\cos2\beta)~+~v_1\,(2\lambda^2-g^2)\,\sin2\beta\right)~,
\]
\[
\lambda_{s_2h_2^+h_2^-} ~=~\frac{1}{2\sq2}\,\left(
v_2\,(g^2+g^{\prime\,2}\cos2\beta)~-~v_1\,(2\lambda^2-g^2)\,\sin2\beta\right)~,
\]
\[
\lambda_{s_2h_1^+h_2^-} ~=~\frac{1}{2\sq2}\,\left(
v_2\,g^{\prime\,2}\sin2\beta~+~v_1\,(2\lambda^2-g^2)\,\cos2\beta\right)~,
\]
\[
\lambda_{s_3h_1^+h_1^-} ~=~\frac{\lambda}{\sq2}\,\left(
2\lambda\,\vs-(\Al + 2 \kappa\,\vs)\,\sin2\beta\right)~,~~~~
\lambda_{s_3h_2^+h_2^-} ~=~\frac{\lambda}{\sq2}\,\left(
2\lambda\,\vs+(\Al + 2 \kappa\,\vs)\,\sin2\beta\right)~,
\]
\[
\lambda_{s_3h_1^+h_2^-} ~=~-\frac{\lambda}{\sq2}\,
(\Al + 2 \kappa\,\vs)\,\cos2\beta~,
\]
\be
\label{chatrilinear}
\lambda_{p_1h_1^+h_2^-} ~=~ \frac{v_2}{2\sq2}\,(2\lambda^2-g^2)~,~~~~
\lambda_{p_2h_1^+h_2^-} ~=~ \frac{v_1}{2\sq2}\,(2\lambda^2-g^2)~,~~~~
\lambda_{p_3h_1^+h_2^-} ~=~ \frac{\lambda}{\sq2}\,(\Al-2\kappa\,\vs)~.
\ee
All of the couplings that cannot be obtained by permuting the indices
of the couplings in eqs.~(\ref{chaquartic}) and (\ref{chatrilinear})
vanish.

\paragraph{Higgs--neutralino couplings:}

In the formalism of two component spinors, the terms in the NMSSM
Lagrangian relevant to the calculation of the neutralino contributions
to the Higgs self energies can be written as
\be
\label{higgsneut}
{\cal L} ~\supset~ 
-\sum_{ik\ell} \lambda_{s_i \psi^0_k\psi^0_\ell} 
~S_i \,\psi^0_k \psi^0_\ell
~-~i\,\sum_{ik\ell} \lambda_{p_i \psi^0_k \psi^0_\ell} 
~P_i\, \psi^0_k \psi^0_\ell~~ +~ {\rm h.c.}~,
\ee
where $\psi^0_i = (-i \tilde b,\,-i\tilde w^0,\, \tilde
h_1^0,\,\tilde h_2^0,\,\tilde s)$ are the neutralino interaction
eigenstates. The couplings in eq.~(\ref{higgsneut}) are symmetric with
respect to the exchange of the neutralino indices $k$ and $\ell$, and
read
\[
\lambda_{s_1 \psi^0_1\psi^0_3} ~=~ 
-\lambda_{s_2 \psi^0_1\psi^0_4} ~=~
-\lambda_{p_1 \psi^0_1\psi^0_3} ~=~
\lambda_{p_2 \psi^0_1\psi^0_4} ~=~
-\frac{\,g^\prime}{4}~,
\]
\[
\lambda_{s_1 \psi^0_2\psi^0_3} ~=~ 
-\lambda_{s_2 \psi^0_2\psi^0_4} ~=~
-\lambda_{p_1 \psi^0_2\psi^0_3} ~=~
\lambda_{p_2 \psi^0_2\psi^0_4} ~=~
\frac{g}{4}~,~~~~~~~
\lambda_{s_3 \psi^0_5\psi^0_5} ~=~ \lambda_{p_3 \psi^0_5\psi^0_5} 
~=~\frac{\kappa}{\sq2}~,
\]
\be
\label{couplneut}
\lambda_{s_1 \psi^0_4\psi^0_5} ~=~ 
\lambda_{s_2 \psi^0_3\psi^0_5} ~=~ 
\lambda_{s_3 \psi^0_3\psi^0_4} ~=~ 
\lambda_{p_1 \psi^0_4\psi^0_5} ~=~ 
\lambda_{p_2 \psi^0_3\psi^0_5} ~=~ 
\lambda_{p_3 \psi^0_3\psi^0_4} ~=~ 
-\frac{\lambda}{2\sq2}~.
\ee
All of the couplings that cannot be obtained by permuting the
neutralino indices of the couplings in the equation above vanish.
The couplings of the Higgs bosons to the neutralino mass eigenstates
$\chi_i^0$ are related to the corresponding couplings to the
neutralino interaction eigenstates as follows
\be
\label{rotneut}
\lambda_{\phi_i\chi^0_k\chi^0_\ell} = 
N^*_{ka}\,N^*_{\ell b}
\,\lambda_{\phi_i\psi^0_a\psi^0_b}~,
\ee
where $\phi_i$ represents either $s_i$ or $p_i$, $N$ is the rotation
matrix defined in eq.~(\ref{diagneut}) and summation over repeated
indices is understood. In the absence of CP-violating phases in the
neutralino mass matrix, the couplings can be taken as real if we allow
for negative neutralino masses.

\paragraph{Higgs--chargino couplings:}

The terms in the NMSSM Lagrangian relevant to the calculation of the
chargino contributions to the Higgs self energies can be written as
\be
\label{higgscha}
{\cal L} ~\supset~ 
-\sum_{ik\ell} \lambda_{s_i \psi^+_k\psi^-_\ell} 
~S_i \,\psi^+_k \psi^-_\ell
~-~i\,\sum_{ik\ell} \lambda_{p_i \psi^+_k \psi^-_\ell} 
~P_i\, \psi^+_k \psi^-_\ell~~ +~ {\rm h.c.}~,
\ee
where $\psi^+_i = (-i\tilde w^+,\, \,\tilde h_2^+)$ and $\psi^-_i =
(-i\tilde w^-,\, \,\tilde h_1^-)$ are the positive and negative
chargino interaction eigenstates, respectively, in the formalism of
two-component spinors. The only non-zero couplings in
eq.~(\ref{higgscha}) are
\[
\lambda_{s_1 \psi^+_1\psi^-_2} ~=~ 
\lambda_{s_2 \psi^+_2\psi^-_1} ~=~
-\lambda_{p_1 \psi^+_1\psi^-_2} ~=~
-\lambda_{p_2 \psi^+_2\psi^-_1} ~=~
\frac{g}{\sq2}~,
\]
\be
\label{couplcha}
\lambda_{s_3 \psi^+_2\psi^-_2} ~=~ 
\lambda_{p_3 \psi^+_2\psi^-_2} ~=~ 
\frac{\lambda}{\sq2}~.
\ee

The mass matrix ${\cal M}_\chi$ for the charginos can be diagonalized
by a biunitary transformation
\be
{\rm diag}(m_{\chi^+_i}) = U^*\,{\cal M}_\chi V^\dagger~,
\ee
where the unitary matrices $U$ and $V$ rotate the negative and
positive chargino interaction eigenstates, respectively, into the
corresponding mass eigenstates
\be
\label{UandV}
\chi^-_i ~=~ U_{ij}\,\psi^-_j~,~~~~~~~
\chi^+_i ~=~ V_{ij}\,\psi^+_j~.
\ee
The couplings of the Higgs bosons to the chargino mass eigenstates are
related to the corresponding couplings to the chargino interaction
eigenstates as follows
\be
\label{rotcha}
\lambda_{\phi_i\chi^+_k\chi^-_\ell} = 
V^*_{ka}\,U^*_{\ell b}
\,\lambda_{\phi_i\psi^+_a\psi^-_b}~,
\ee
where $\phi_i$ represents either $s_i$ or $p_i$ and summation over
repeated indices is understood. In the absence of CP-violating phases
in the chargino mass matrix, the couplings can be taken as real if we
allow for negative chargino masses.

\paragraph{$Z$--neutralino and $Z$--chargino couplings:}

Since the singlino is neutral with respect to the MSSM gauge sector,
the couplings of the $Z$ boson to the neutralinos (and, of course, to
the charginos) in the interaction basis are the same as in the MSSM.
In the formalism of two-component spinors they read
\be
\label{Zlagr}
{\cal L} ~\supset~ -\sum_{ij}\left(
\lambda_{Z\psi^0_i\psi^0_j}\,Z_\mu\,\bar{\psi}^0_i\bar\sigma^\mu\psi^0_j ~+~
{\rm h.c.}\right)
~-~ \sum_{ij}\lambda_{Z\psi^+_i\psi^+_j}\,Z_\mu\,\bar{\psi}^+_i\bar\sigma^\mu\psi^+_j
~-~ \sum_{ij}\lambda_{Z\psi^-_i\psi^-_j}\,Z_\mu\,\psi^-_i\sigma^\mu\bar{\psi}^-_j~.
\ee
The only non-zero couplings are
\be
\label{Zcoupl}
\lambda_{Z\psi^0_3\psi^0_3}~=\, - \lambda_{Z\psi^0_4\psi^0_4}~=~
\frac{\bar g}{2\sq2}~,~~~~
\lambda_{Z\psi^+_1\psi^+_1}~=\, \lambda_{Z\psi^-_1\psi^-_1}~=~
\frac{g^2}{\sq2\,\bar g}~,~~~~
\lambda_{Z\psi^+_2\psi^+_2}~=\, \lambda_{Z\psi^-_2\psi^-_2}~=~
\frac{\bar g}{\sq2}\,\cos 2 \theta_\smallw~.
\ee
In terms of the neutralino and chargino mass eigenstates, the couplings read
\be
\label{Zcouplrot}
\lambda_{Z\chi^0_i\chi^0_j} = N_{ik}N^*_{j\ell}\,\lambda_{Z\psi^0_k\psi^0_\ell}~,~~~~~~
\lambda_{Z\chi^+_i\chi^+_j} = V_{ik}V^*_{j\ell}\,\lambda_{Z\psi^+_k\psi^+_\ell}~,~~~~~~
\lambda_{Z\chi^-_i\chi^-_j} = U^*_{ik}U_{j\ell}\,\lambda_{Z\psi^-_k\psi^-_\ell}~.
\ee

\paragraph{$W$--neutralino--chargino couplings:}

The couplings of the $W$ boson with charginos and neutralinos are
the same as in the MSSM, but we give them for completeness. In the
formalism of two-component spinors, they read
\be
\label{Wlagr}
{\cal L} ~\supset~ -\sum_{ij}
\lambda_{W\psi^0_i\psi^+_j}\,W^+_\mu\,\bar{\psi}^+_j\bar\sigma^\mu\psi^0_i 
~- \sum_{ij}\lambda_{Z\psi^0_i\psi^-_j}\,W^+_\mu\,\psi^-_j\sigma^\mu\bar\psi^0_i
~+~ {\rm h.c.}
\ee
The only non-zero couplings are
\be
\label{Wcoupl}
\lambda_{W\psi^0_2\psi^+_1}~=\, \lambda_{W\psi^0_2\psi^-_1}~=~ -\,g~,~~~~
\lambda_{W\psi^0_4\psi^+_2}~=\, -\lambda_{W\psi^0_3\psi^-_2}~=~ \frac{g}{\sq2}~,~~~~
\ee
In terms of the neutralino and chargino mass eigenstates, the couplings read
\be
\label{Wcouplrot}
\lambda_{W\chi^0_i\chi^+_j} = N^*_{ik}V_{j\ell}\,\lambda_{W\psi^0_k\psi^+_\ell}~,~~~~~~
\lambda_{W\chi^0_i\chi^-_j} = N_{ik}U^*_{j\ell}\,\lambda_{Z\psi^0_k\psi^-_\ell}~.
\ee

\end{appendletterA}

\section*{Appendix B: one-loop self energies and tadpoles}

\begin{appendletterB}

In this appendix we list the explicit formulae for the one-loop self
energies and tadpole diagrams that are necessary to the calculation of
the neutral Higgs boson masses. The calculation is performed in the 't
Hooft-Feynman gauge, in which the Goldstone bosons and the ghosts have
the same masses as the corresponding gauge bosons.

\paragraph{Scalar self energies} The contributions to the scalar self 
energies from matter-fermion loops and from loops involving gauge
bosons or ghosts are essentially the same as in the MSSM, and
read~\cite{pbmz}
\bea
16\,\pi^2\,\Pi_{s_1s_1}^{f,\,V}(p^2) &=&
3\,h_b^2\,\left[(p^2-4\,\mb^2)\,B_0(\mb,\mb) - 2\,A_0(\mb)\right] \nn\\
&+&
~\;h_\tau^2\,\left[(p^2-4\,m_\tau^2)\,B_0(m_\tau,m_\tau) - 2\,A_0(m_\tau)\right] 
\nn\\&+& 
\frac{g^2}{2}\,\sum_{i=1}^2\,\left(R^\smallC_{i1}\right)^2\,F(m_{h^\pm_i},M_W)
~+~\frac{\gbq}{2}\,\sum_{i=1}^3\,\left(R^\smallP_{i1}\right)^2\,F(m_{a_i},M_Z)
\nn\\
&+&\frac{7}{2}\,c_\beta^2\,\left[ g^2\,M_W^2\,B_0(M_W,M_W) +
\gbq \,M_Z^2\,B_0(M_Z,M_Z)\right]\nn\\
&+& 2\,g^2 A_0(M_W) + 2\, \gbq A_0(M_Z)~,\\
\nn\\
16\,\pi^2\,\Pi_{s_2s_2}^{f,\,V}(p^2) &=&
3\,h_t^2\,\left[(p^2-4\,\mt^2)\,B_0(\mt,\mt) - 2\,A_0(\mt)\right] \nn\\
&+& 
\frac{g^2}{2}\,\sum_{i=1}^2\,\left(R^\smallC_{i2}\right)^2\,F(m_{h^\pm_i},M_W)
~+~\frac{\gbq}{2}\,\sum_{i=1}^3\,\left(R^\smallP_{i2}\right)^2\,F(m_{a_i},M_Z)
\nn\\
&+&\frac{7}{2}\,s_\beta^2\,\left[ g^2\,M_W^2\,B_0(M_W,M_W) +
\gbq \,M_Z^2\,B_0(M_Z,M_Z)\right]\nn\\
&+& 2\,g^2 A_0(M_W) + 2\, \gbq A_0(M_Z)~,\\
\nn\\
16\,\pi^2\,\Pi_{s_1s_2}^{f,\,V}(p^2) &=&
-\frac{g^2}{2}\,\sum_{i=1}^2\,R^\smallC_{i1}R^\smallC_{i2}\,F(m_{h^\pm_i},M_W)
~-~\frac{\gbq}{2}\,\sum_{i=1}^3\,R^\smallP_{i1}R^\smallP_{i2}
\,F(m_{a_i},M_Z)
\nn\\
&+&\frac{7}{2}\,s_\beta\,c_\beta\,\left[ g^2\,M_W^2\,B_0(M_W,M_W) +
\gbq \,M_Z^2\,B_0(M_Z,M_Z)\right]~,
\eea
where we neglected the Yukawa couplings of the first two
generations. The loop functions $A_0(m)$, $B_0(m_1,m_2)$ and
$F(m_1,m_2)$ are defined in appendix B of ref.~\cite{pbmz}, and they
depend also on the external momentum $p^2$ and on the renormalization
scale $Q$. Here and in the following we use for brevity the notation
$c_\phi \equiv \cos\phi\,,~s_\phi \equiv \sin\phi$, for a generic
angle $\phi$. Also, $h_i^\pm$ stands for ($G^\pm,H^\pm$), $a_i$ stands
for ($G^0,A_1,A_2$), and the rotation matrices $R^\smallC$ and
$R^\smallP$ are defined in eqs.~(\ref{rothpm}) and (\ref{rotodd}),
respectively. The contributions to the self energies involving the
field $S_3$ are zero because the singlet does not couple to the gauge
sector nor to the matter fermions.

The sfermion contributions to the scalar self energies read
\be
16\,\pi^2\,\Pi_{s_is_j}^{\tilde f}(p^2) ~=~
\sum_{\tilde f} \,\sum_{k=1}^{n_{\tilde f}}\,2\,N_c^f
\,\lambda_{s_is_j\tilde f_k\tilde f_k}\,A_0(m_{\tilde f_k}) ~+~
\sum_{\tilde f} \,\sum_{k,\ell=1}^{n_{\tilde f}}\,
N_c^f\,\lambda_{s_i\tilde f_k\tilde f_\ell}
\,\lambda_{s_j\tilde f_\ell\tilde f_k}
\,B_0(m_{\tilde f_k},m_{\tilde f_\ell})~,
\ee
where the first sum in each term runs over all the sfermion species
and the second over the sfermion mass eigenstates. $N_c^f$ is the
number of colours for the sfermions $\tilde f$, while $n_{\tilde f}$ is 1 for
the sneutrinos and 2 for all of the other sfermions. The quartic and
trilinear Higgs--sfermion couplings are obtained by combining
eqs.~(\ref{hsfquau})--(\ref{rotsf}).

The Higgs contributions to the scalar self energies read
\bea
16\,\pi^2\,\Pi_{s_is_j}^{H}(p^2) &=&
\sum_{k=1}^3 \,2\,\lambda_{s_is_j h_k h_k}
\,A_0(m_{h_k}) +
\sum_{k,\ell=1}^3 \,2\,\lambda_{s_i h_k h_\ell}\,\lambda_{s_j h_k h_\ell}
\,B_0(m_{h_k},m_{h_\ell})\nn\\
\nn\\
&+&
\sum_{k=1}^3 \,2\,\lambda_{s_is_j a_k a_k}
\,A_0(m_{a_k}) +
\sum_{k,\ell=1}^3 \,2\,\lambda_{s_i a_k a_\ell}\,\lambda_{s_j a_k a_\ell}
\,B_0(m_{a_k},m_{a_\ell})\nn\\
&+& \sum_{n=1}^2 \,2\,\lambda_{s_is_j h^+_k h^-_k}
\,A_0(m_{h^\pm_k}) ~+
\sum_{m,n=1}^2 \lambda_{s_i h^+_k h^-_\ell}\,\lambda_{s_j h^+_\ell h^-_k}
\,B_0(m_{h^\pm_k},m_{h^\pm_\ell})~,
\eea
where the neutral couplings $\lambda_{s_i s_j \phi_k \phi_\ell}$ and
$\lambda_{s_i \phi_k \phi_\ell}$, where $\phi$ represents either $h$
or $a$, are obtained by combining
eqs.~(\ref{selfquartic})--(\ref{selfscalrot}), while the charged
couplings $\lambda_{s_i s_j h^+_k h^-_\ell}$ and $\lambda_{s_i h^+_k
  h^-_\ell}$ are given in eqs.~(\ref{chaquartic}) and
(\ref{chatrilinear}), respectively.

Finally, the chargino and neutralino contributions to the scalar self
energies read
\bea
16\,\pi^2\,\Pi_{s_is_j}^{\chi}(p^2) &=&
4\,\sum_{k,\ell=1}^5 \,\biggr[
{\rm Re}(\lambda^*_{s_i\chi^0_k \chi^0_\ell}\,\lambda_{s_j\chi^0_k \chi^0_\ell})\,
G(m_{\chi^0_k},m_{\chi^0_\ell}) \nn\\
&&~~~~~~~~~~~-2\,m_{\chi^0_k}\,m_{\chi^0_\ell}\,
{\rm Re}(\lambda_{s_i\chi^0_k \chi^0_\ell}\,\lambda_{s_j\chi^0_k \chi^0_\ell})\,
B_0(m_{\chi^0_k},m_{\chi^0_\ell})\biggr] \nn\\
&+&
2\,\sum_{k,\ell=1}^2 \,\biggr[
{\rm Re}(\lambda^*_{s_i\chi^+_k \chi^-_\ell}\,\lambda_{s_j\chi^+_k \chi^-_\ell})\,
G(m_{\chi^\pm_k},m_{\chi^\pm_\ell}) \nn\\
&&~~~~~~~~~~~-2\,m_{\chi^\pm_k}\,m_{\chi^\pm_\ell}\,
{\rm Re}(\lambda_{s_i\chi^+_k \chi^-_\ell}\,\lambda_{s_j\chi^+_\ell \chi^-_k})\,
B_0(m_{\chi^\pm_k},m_{\chi^\pm_\ell})\biggr]~,
\eea
where the loop function $G(m_1,m_2)$ is defined in appendix B of
ref.~\cite{pbmz}, the Higgs-neutralino couplings are obtained by
combining eqs.~(\ref{couplneut}) and (\ref{rotneut}), and the
Higgs-chargino couplings are obtained by combining
eqs.~(\ref{couplcha}) and (\ref{rotcha}).

\paragraph{Pseudoscalar self energies} The contributions to the 
pseudoscalar self energies from matter-fermion loops and from loops
involving gauge bosons or ghosts read
\bea
16\,\pi^2\,\Pi_{p_1p_1}^{f,\,V}(p^2) &=&
3\,h_b^2\,\left[p^2\,B_0(\mb,\mb) - 2\,A_0(\mb)\right] ~+~
h_\tau^2\,\left[p^2\,B_0(m_\tau,m_\tau) - 2\,A_0(m_\tau)\right] \nn\\
&+& 
\frac{g^2}{2}\,\sum_{i=1}^2\,\left(R^\smallC_{i1}\right)^2\,F(m_{h^\pm_i},M_W)
~+~\frac{\gbq}{2}\,\sum_{i=1}^3\,\left(R^\smallS_{i1}\right)^2\,F(m_{h_i},M_Z)
\nn\\
&+& \frac{g^2}{2}\,c_\beta^2 \,M_W^2\,B_0(M_W,M_W)
+ 2\,g^2 A_0(M_W) + 2\, \gbq A_0(M_Z)~,\\
\nn\\
16\,\pi^2\,\Pi_{p_2p_2}^{f,\,V}(p^2) &=&
3\,h_t^2\,\left[p^2\,B_0(\mt,\mt) - 2\,A_0(\mt)\right] \nn\\
&+& 
\frac{g^2}{2}\,\sum_{i=1}^2\,\left(R^\smallC_{i2}\right)^2\,F(m_{h^\pm_i},M_W)
~+~\frac{\gbq}{2}\,\sum_{i=1}^3\,\left(R^\smallS_{i2}\right)^2\,F(m_{h_i},M_Z)
\nn\\
&+&  \frac{g^2}{2}\,s_\beta^2 \,M_W^2\,B_0(M_W,M_W)
+ 2\,g^2 A_0(M_W) + 2\, \gbq A_0(M_Z)~,\\
\nn\\
16\,\pi^2\,\Pi_{p_1p_2}^{f,\,V}(p^2) &=&
\frac{g^2}{2}\,\sum_{i=1}^2\,R^\smallC_{i1}R^\smallC_{i2}\,F(m_{h^\pm_i},M_W)
-\frac{\gbq}{2}\,\sum_{i=1}^3\,R^\smallS_{i1}R^\smallS_{i2}
\,F(m_{h_i},M_Z)\nn\\
&-& \frac{g^2}{2}\,c_\beta\,s_\beta \,M_W^2\,B_0(M_W,M_W)~,
\eea
the rotation matrices $R^\smallC$ and $R^\smallS$ are defined in
eqs.~(\ref{rothpm}) and (\ref{roteven}), respectively.  The
contributions to the self energies involving the field $P_3$ are zero
because the singlet does not couple to the gauge sector nor to the
matter fermions.

The sfermion contributions to the pseudoscalar self energies read
\be
16\,\pi^2\,\Pi_{p_ip_j}^{\tilde f}(p^2) ~=~
\sum_{\tilde f} \,\sum_{k=1}^{n_{\tilde f}}\,2\,N_c^f
\,\lambda_{p_ip_j\tilde f_k\tilde f_k}\,A_0(m_{\tilde f_k}) ~-~
\sum_{\tilde f} \,\sum_{k,\ell=1}^{n_{\tilde f}}\,N_c^f\,
\lambda_{p_i\tilde f_k\tilde f_\ell}
\,\lambda_{p_j\tilde f_\ell\tilde f_k}
\,B_0(m_{\tilde f_k},m_{\tilde f_\ell})~,
\ee
where the quartic and trilinear Higgs--sfermion couplings are obtained
by combining eqs.~(\ref{hsfquau})--(\ref{rotsf}).

The Higgs contributions to the pseudoscalar self energies read
\bea
16\,\pi^2\,\Pi_{p_ip_j}^{H}(p^2) &=&
\sum_{k=1}^3 \,2\,\lambda_{p_ip_j h_k h_k}
\,A_0(m_{h_k}) +
\sum_{k=1}^3 \,2\,\lambda_{p_ip_j a_k a_k}
\,A_0(m_{a_k}) \nn\\
&+&
\sum_{k,\ell=1}^3 \,\lambda_{p_i a_k h_\ell}\,\lambda_{p_j a_k h_\ell}
\,B_0(m_{a_k},m_{h_\ell})\nn\\
&+& \sum_{n=1}^2 \,2\,\lambda_{p_ip_j h^+_k h^-_k}
\,A_0(m_{h^\pm_k}) ~-
\sum_{m,n=1}^2 \lambda_{p_i h^+_k h^-_\ell}\,\lambda_{p_j h^+_\ell h^-_k}
\,B_0(m_{h^\pm_k},m_{h^\pm_\ell})~,
\eea
where the neutral couplings $\lambda_{p_i p_j h_k h_\ell}$,
$\lambda_{p_i p_j a_k a_\ell}$ and $\lambda_{p_i a_k h_\ell}$ are
obtained by combining eqs.~(\ref{selfquartic}) and
(\ref{selftrilinear}) with eq.~(\ref{selfpseurot}), while the charged
couplings $\lambda_{p_i p_j h^+_k h^-_\ell}$ and $\lambda_{p_i h^+_k
  h^-_\ell}$ are given in eqs.~(\ref{chaquartic}) and
(\ref{chatrilinear}), respectively.

Finally, the chargino and neutralino contributions to the pseudoscalar
self energies read
\bea
16\,\pi^2\,\Pi_{p_ip_j}^{\chi}(p^2) &=&
4\,\sum_{k,\ell=1}^5 \,\biggr[
{\rm Re}(\lambda^*_{p_i\chi^0_k \chi^0_\ell}\,\lambda_{p_j\chi^0_k \chi^0_\ell})\,
G(m_{\chi^0_k},m_{\chi^0_\ell}) \nn\\
&&~~~~~~~~~~~+2\,m_{\chi^0_k}\,m_{\chi^0_\ell}\,
{\rm Re}(\lambda_{p_i\chi^0_k \chi^0_\ell}\,\lambda_{p_j\chi^0_k \chi^0_\ell})\,
B_0(m_{\chi^0_k},m_{\chi^0_\ell})\biggr] \nn\\
&+&
2\,\sum_{k,\ell=1}^2 \,\biggr[
{\rm Re}(\lambda^*_{p_i\chi^+_k \chi^-_\ell}\,\lambda_{p_j\chi^+_k \chi^-_\ell})\,
G(m_{\chi^\pm_k},m_{\chi^\pm_\ell}) \nn\\
&&~~~~~~~~~~~+2\,m_{\chi^\pm_k}\,m_{\chi^\pm_\ell}\,
{\rm Re}(\lambda_{p_i\chi^+_k \chi^-_\ell}\,\lambda_{p_j\chi^+_\ell \chi^-_k})\,
B_0(m_{\chi^\pm_k},m_{\chi^\pm_\ell})\biggr]~,
\eea
where the Higgs-neutralino couplings are obtained by combining
eqs.~(\ref{couplneut}) and (\ref{rotneut}), and the Higgs-chargino
couplings are obtained by combining eqs.~(\ref{couplcha}) and
(\ref{rotcha}).

\paragraph{Scalar tadpoles}

The contributions to the scalar tadpoles from matter-fermion loops and
from loops involving gauge bosons or ghosts are the same as in the
MSSM, and read~\cite{pbmz}
\bea
16\,\pi^2\,T_1^{\,f,\,V} &=&
-6\,\sq2\,h_b\,m_b\,A_0(m_b) -2\,\sq2\,h_\tau\,m_\tau\,A_0(m_\tau)
+\frac{3\,v_1}{\sq2}\left(g^2\,A_0(M_W) + \gbq \,A_0(M_Z)\right)~,\nn\\
&&\\
16\,\pi^2\,T_2^{\,f,\,V} &=&
-6\,\sq2\,h_t\,m_t\,A_0(m_t) 
+\frac{3\,v_2}{\sq2}\left(g^2\,A_0(M_W) + \gbq \,A_0(M_Z)\right)~.
\eea
The contributions to $T_3$ are zero because the singlet does not
couple to the gauge sector nor to the matter fermions.

The contributions to the tadpoles from loops involving sfermions or
Higgs bosons read
\be
16\,\pi^2\,T_i^{\tilde f\,,\phi} ~=~
\sum_{\tilde f} \,\sum_{k=1}^{n_{\tilde f}}\,N_c^f
\,\lambda_{s_i\tilde f_k\tilde f_k}\,A_0(m_{\tilde f_k}) ~+~
\sum_\phi \,\sum_{k=1}^{n_\phi}\,
\,\lambda_{s_i\phi_k\phi_k}\,A_0(m_{\tilde \phi_k})~,
\ee
where the first sum in the second term runs over all the Higgs bosons
(i.e.,~$\phi = h,a,h^\pm$), and $n_\phi$ is 2 for $h^\pm$ and 3 for
$h$ and $a$. The trilinear Higgs--sfermion couplings are obtained by
combining eqs.~(\ref{hsftriu})--(\ref{rotsf}); the trilinear
neutral-Higgs self-couplings are obtained by combining
eqs.~(\ref{selftrilinear}) and (\ref{selfscalrot}); the trilinear
couplings with the charged Higgs bosons are given in
eq.~(\ref{chatrilinear}).

Finally, the contributions to the tadpoles from loops involving neutralinos
or charginos read
\be
16\,\pi^2\,T_i^{\chi} ~=~
- 4\,\sum_{k=1}^{5}\,\lambda_{s_i\chi^0_k\chi^0_k}\,m_{\chi^0_k}\,A_0(m_{\chi^0_k})
~- 4\,\sum_{k=1}^{2}\,\lambda_{s_i\chi^+_k\chi^-_k}\,m_{\chi^\pm_k}\,A_0(m_{\chi^\pm_k})~,
\ee
where the Higgs-neutralino couplings are obtained by combining
eqs.~(\ref{couplneut}) and (\ref{rotneut}), and the Higgs-chargino
couplings are obtained by combining eqs.~(\ref{couplcha}) and
(\ref{rotcha}).

\paragraph{Vector boson self energies}

Since the singlet superfield is neutral under the MSSM gauge group,
the vector boson self energies in the NMSSM differ from the
corresponding MSSM quantities only through the effect of the mixing of
the singlet and singlino with the MSSM Higgs bosons and higgsinos,
respectively. The transverse parts of the $W$ and $Z$ self energies
appearing in eqs.~(\ref{gauge1}) and (\ref{gauge2}) can be obtained by
inserting in eqs.~(D.4) and (D.9) of ref.~\cite{pbmz} appropriate
combinations of the Higgs scalar and pseudoscalar mixing matrices, as
well as the chargino and neutralino couplings defined in appendix A.

The transverse part of the $Z$-boson self energy reads
\bea
16\,\pi^2\,\Pi^T_{ZZ}(p^2) &=&
2\,\gbq\,M_Z^2\,\sum_{i=1}^3\,
\left(R^\smallS_{i1}\,c_\beta+R^\smallS_{i2}\,s_\beta\right)^2\,
B_{0}(m_{h_i},M_Z)\nn\\
& - & 2\,\gbq\,\sum_{i,j=1}^3\,\left(R^\smallS_{i1}\,R^\smallP_{j1}
-R^\smallS_{i2}\,R^\smallP_{j2}\right)^2\,
\widetilde B_{22}(m_{h_i},m_{a_j})
~-~2\,\gbq\,c_{2\theta_\smallw}^2\,\sum_{i}^2 
\widetilde B_{22}(m_{h^\pm_i},m_{h^\pm_i})\nn\\
&-&4\,\gbq\,c_{\theta_\smallw}^4\,\left(2\,p^2+M_W^2-M_Z^2\,
\frac{s_{\theta_\smallw}^4}{c_{\theta_\smallw}^2}\right)\,B_0(M_W,M_W) 
~-~ 16\,\gbq\,c_{\theta_\smallw}^4\,\widetilde B_{22}(M_W,M_W)\nn\\
&+&2\,\gbq\,\sum_{f}\,N_c^f\,\left[
\left(g_{f_L}^2+g_{f_R}^2\right)\,H(m_f,m_f)-4\,g_{f_L}\,g_{f_R}\,m_f^2\,
B_0(m_f,m_f)\right]\nn\\
&-&8\,\gbq\,\sum_{\tilde f}\sum_{i,j=1}^{n_{\tilde f}} \,N_c^f\,
\left(g_{f_L}\,R^{\tilde f}_{i1}\,R^{\tilde f}_{j1}
-g_{f_R}\,R^{\tilde f}_{i2}\,R^{\tilde f}_{j2}\right)^2\,
\widetilde B_{22}(m_{\tilde f_i},m_{\tilde f_j})\nn\\
&+& 4\,\sum_{i,j=1}^5\,|\lambda_{Z\chi^0_i\chi^0_j}|^2\,
\left[H(m_{\chi^0_i},m_{\chi^0_j})-2\,m_{\chi^0_i}\,m_{\chi^0_j}\,
B_0(m_{\chi^0_i},m_{\chi^0_j})\right]\nn\\
&+& \sum_{i,j=1}^2\,\biggr[\left(|\lambda_{Z\chi^+_i\chi^+_j}|^2
+|\lambda_{Z\chi^-_i\chi^-_j}|^2\right)\,H(m_{\chi^\pm_i},m_{\chi^\pm_j})\nn\\
&&~~~~~~~~~~~~~~~~~~~~+4\,m_{\chi^\pm_i}\,m_{\chi^\pm_j}\,{\rm Re}\left(
\lambda_{Z\chi^+_i\chi^+_j}^*\lambda_{Z\chi^-_i\chi^-_j}\right)\,
B_0(m_{\chi^\pm_i},m_{\chi^\pm_j})\biggr]~,
\eea
where the loop functions $\widetilde B_{22}(m_1,m_2)$ and $H(m_1,m_2)$ are
defined in appendix B of ref.~\cite{pbmz}, $g_f$ is defined after
eq.~(\ref{hsftrid}), and the couplings $\lambda_{Z\chi^0_i\chi^0_j}$,
$\,\lambda_{Z\chi^+_i\chi^+_j}$ and $\lambda_{Z\chi^-_i\chi^-_j}$ are
obtained by combining eqs.~(\ref{Zcoupl}) and (\ref{Zcouplrot}).

Finally, the transverse part of the $W$-boson self energy reads
\bea
16\,\pi^2\,\Pi^T_{WW}(p^2) &=&
g^2\,M_W^2\,\sum_{i=1}^3\,
\left(R^\smallS_{i1}\,c_\beta+R^\smallS_{i2}\,s_\beta\right)^2\,
B_{0}(m_{h_i},M_W)\nn\\
&-& g^2\,\sum_{i=1}^3\sum_{j=1}^2 \left(R^\smallS_{i1}\,R^\smallC_{j1}
-R^\smallS_{i2}\,R^\smallC_{j2}\right)^2\,
\widetilde B_{22}(m_{h_i},m_{h^\pm_j}) \nn\\
&-& g^2\,\sum_{i=1}^3 \sum_{j=1}^2 \left(R^\smallP_{i1}\,R^\smallC_{j1}
+R^\smallP_{i2}\,R^\smallC_{j2}\right)^2 
\,\widetilde B_{22}(m_{a_i},m_{h^\pm_j})\nn\\
&-& g^2\,\left[\left(4\,p^2+M_W^2+M_Z^2\right)\,c_{\theta_\smallw}^2
- M_Z^2\,s_{\theta_\smallw}^4\right]\,B_0(M_Z,M_W)\nn\\
&-&8\,g^2\,c_{\theta_\smallw}^2\,\widetilde B_{22}(M_Z,M_W)
~-~ g^2\,s_{\theta_\smallw}^2\,\left[
8\,\widetilde B_{22}(M_W,0) + 4\,p^2\,B_0(M_W,0)\right]\nn\\
&+&\frac{3\, g^2}{2}\,\sum_{u/d}\,H(m_u,m_d)
~+~\frac{g^2}{2}\,\sum_{e/\nu}\,H(m_e,0)\nn\\
&-&6\,g^2\,\sum_{\tilde u/\tilde d}\,\sum_{i,j=1}^{2} \,
\left(R^{\,\tilde u}_{i1}\,R^{\,\tilde d}_{j1}\right)^2\,
\widetilde B_{22}(m_{\tilde u_i},m_{\tilde d_j})
~-~2\,g^2\,\sum_{\tilde e/\tilde \nu}\,\sum_{i=1}^{2} \,
\left(R^{\,\tilde e}_{i1}\right)^2\,
\widetilde B_{22}(m_{\tilde e_i},m_{\tilde \nu})\nn\\
&+& \sum_{i=1}^5\,\sum_{j=1}^2\,\biggr[\left(|\lambda_{W\chi^0_i\chi^+_j}|^2
+|\lambda_{W\chi^0_i\chi^-_j}|^2\right)\,H(m_{\chi^0_i},m_{\chi^\pm_j})\nn\\
&&~~~~~~~~~~~~~~~~~~~~+4\,m_{\chi^0_i}\,m_{\chi^\pm_j}\,{\rm Re}\left(
\lambda_{W\chi^0_i\chi^+_j}^*\lambda_{W\chi^0_i\chi^-_j}\right)\,
B_0(m_{\chi^0_i},m_{\chi^\pm_j})\biggr]~,
\eea
where the sums in the fermion contributions and the first sums in the
sfermion contributions run over the three families of (s)quarks and
(s)leptons, and the couplings $\lambda_{W\chi^0_i\chi^+_j}$ and
$\lambda_{W\chi^0_i\chi^-_j}$ are obtained by combining
eqs.~(\ref{Wcoupl}) and (\ref{Wcouplrot}).

\end{appendletterB}


\section*{Appendix C: derivatives of the two-loop effective potential}
\begin{appendletterC}

We provide in this appendix the explicit formulae for the derivatives
of the two-loop contribution to the effective potential involving top,
stop, gluon and gluino. In units of $\as\, C_F N_c\,/(4 \pi)^3$, where
$C_F = 4/3$ and $N_c = 3$ are colour factors, and in terms of the
field-dependent quantities defined in
eqs.~(\ref{topstop})--(\ref{phases}), the $\oas$ contribution to
$\Veff$ reads
\bea
\label{v2as}
\Delta V & = & 
2\, J(\t,\t) - 4\,\t\,I(\t,\t,0) + \nn \\
&+& 
\biggr\{ 2\,\tu\,I(\tu,\tu,0) + 2\,L(\tu,\g,\t)
- 4\,m_t\,\mg\,s_{2\bar\theta}\,c_{\varphi - \tilde{\varphi}}
\,I(\tu,\g,\t) \nn\\
&& +\,\frac12\,  (1+c_{2\bar\theta}^2)\,J(\tu,\tu) 
+ \frac{s_{2\bar\theta}^2}{2} J(\tu,\td)\;\; 
+ \;\; \left[ \tul \leftrightarrow \tdl\,,\,
s_{2\bar\theta} \rightarrow - s_{2\bar\theta}\right] \biggr\}\,,
\eea

\noindent
where the functions $I\,, J$ and $L$ are defined in appendix D. The
derivatives of $\Delta V$ that involve only the field-dependent stop
mixing angle $\bar\theta$ and phase difference $\varphi -
\tilde{\varphi}$ can be straightforwardly computed from
eq.~(\ref{v2as}). In units of $\as\, C_F N_c\,/(4 \pi)^3$, they read
\bea
\label{DVcdtq}
\DVcdtq &=& \frac12\,\left[ J(\tu,\tu) + J(\td,\td)\right] 
- J(\tu,\td) \nn\\
&&+ 2\, \frac{\mg\,m_t}{\sdt} \left[I(\tu,\g,\t)-I(\td,\g,\t)\right]~,\\
\DVcdtqcdtq &=& -\frac{z_t}{4\,\sdt^4}\,\DVcptmptt~=~
 \frac{\mg\,m_t}{\sdt^3} \left[ I(\tu,\g,\t) - I(\td,\g,\t)\right]~,
\eea
The explicit expressions for derivatives of $\Delta V$ that involve
the quark or squark masses are somewhat lengthier. In units of $\as\,
C_F N_c\,/(4 \pi)^3$, they read
\bea
\DVtu &=&-6\,\tu + 2\,\mg\,m_t\,\sdt
+4\,\t\,\left(1-\log\frac{\t}{\tu}\right)
+4\,\g\,\left(1-\log\frac{\g}{\tu}\right)\nn\\
&&+\left[\left(5-\cdt^2\right)\,\tu-\sdt^2\,\td
-4\,\mg\,m_t\,\sdt\right]\,\log\frac{\tu}{Q^2}\nn\\
&&+\left(-3+\cdt^2\right)\,\tu\,\log^2\frac{\tu}{Q^2}
+\sdt^2\,\td\,\log\frac{\tu}{Q^2}\,\log\frac{\td}{Q^2}\nn\\
&&-\left[2\,\left(\g+\t-\tu\right)-2\,\mg\,\mt\,\sdt\right]\,
\left(\log\frac{\t}{Q^2}\,\log\frac{\tu}{\g}
+\log\frac{\tu}{Q^2}\,\log\frac{\g}{Q^2}\right)\nn\\
\label{dvdtu}
&&+\left[\frac{2}{\t}\,\left(\Delta +2\,\g\,\t\right)
-\frac{2\,\mg\,\sdt}{\mt}\,\left(\g+\t-\tu\right)\right]\,
\Phi(\tu,\g,\t)~,\\
&&\nn\\
\DVtutu &=&
-\left(1+\cdt^2\right)
+\frac4\tu\,\left(\g+\t-\mg\,\mt\,\sdt\right)
-\sdt^2\,\frac{\td}{\tu}\,\left(1-\log\frac{\td}{Q^2}\right)\nn\\
&&+\left[3+\cdt^2+\frac{8\,\g\,\t}{\Delta}
-\frac{4\,\mg\,\mt\,\sdt}{\Delta}\,\left(\g+\t-\tu\right)\right]
\,\log\frac{\tu}{Q^2}\nn\\
&&-\frac{4\,\t}{\Delta\,\tu}\,
\left[\Delta -\g\,\left(\g-\t-\tu\right)
+\,\mg\,\mt\,\sdt\,\left(\g-\t+\tu\right)\right]\,\log\frac{\t}{Q^2}\nn\\
&&-\frac{4\,\g}{\Delta\,\tu}\,
\left[\Delta +\t\,\left(\g-\t+\tu\right)
-\,\mg\,\mt\,\sdt\,\left(\g-\t-\tu\right)\right]\,\log\frac{\g}{Q^2}\nn\\
&&+\left(-3+\cdt^2\right)\,\log^2\frac{\tu}{Q^2}
+2\,\left(\log\frac{\t}{Q^2}\,\log\frac{\tu}{\g}
+\log\frac{\tu}{Q^2}\,\log\frac{\g}{Q^2}\right)\nn\\
&&-\frac{2}{\Delta\,\t}\,\left[
\left(\g+\t-\tu\right)\,\left(\Delta-2\,\g\,\t\right)
+ 4\,\mg^3\,\mt^3\,\sdt\right]\,\Phi(\tu,\g,\t)~,\\
&&\nn\\
\DVcdtqtu &=&
\left[\td\,\left(1-\log\frac{\td}{Q^2}\right)
-\tu\,\left(1-\log\frac{\tu}{Q^2}\right)\right]\,\log\frac{\tu}{Q^2}\nn\\
&&-\frac{\mg\,\mt}{\sdt}\biggr[1-2\,\log\frac{\tu}{Q^2}+
\log\frac{\tu}{\g}\,\log\frac{\t}{Q^2}+\log\frac{\tu}{Q^2}\,\log\frac{\g}{Q^2}
\nn\\&&~~~~~~~~~~~~~~~~~~~~~~~~~~~~~
-\frac{1}{\t}\left(\g+\t-\tu\right)\,\Phi(\tu,\g,\t)\biggr]~,\\
&&\nn\\
\DVttu &=& \frac{\mg\,\sdt}{\mt}
+\frac{4\,\g}{\Delta}\,\left[\tu-\g-\t+2\,\mt\,\mg\,\sdt\right]\,
\log\frac{\g}{Q^2}\nn\\
&&+\frac{4}{\Delta}\,\left[2\,\g\,\t - \mt\,\mg\,\sdt\,
\left(\g+\t-\tu\right)\right]\,\log\frac{\t}{Q^2}
\nn\\&&+\frac{2}{\Delta}\,\left[
2\,\g\,\left(\g-\t-\tu\right)-\frac{\mg\,\sdt}{\mt}\,
\left(\Delta - 2\,\t\,(\t-\g-\tu)\right)\right]\,\log\frac{\tu}{Q^2}\nn\\
&&+\left(-2+\frac{\mg\,\sdt}{\mt}\right)
\left(\log\frac{\tu}{\g}\,\log\frac{\t}{Q^2}
+\log\frac{\tu}{Q^2}\,\log\frac{\g}{Q^2}\right)\nn\\
&&+\frac{1}{\Delta\,\t}\,\biggr\{
\frac{\mg\,\sdt}{\mt}\left[\Delta\,\left(\tu-\g-3\,\t\right)+2\,\t
\left((\t-\tu)^2-\gq\right)\right]\nn\\
\label{dvdtucdt}
&&+2\,\left(\g-\tu\right)^3 +2\,\t\,\left[\Delta+\left(2\,\tu-\t\right)\,
\left(\g+\tu\right)\right]\biggr\}\,\Phi(\tu,\g,\t)~,\\
&&\nn\\
\DVtutd &=& \sdt^2\,\log\frac{\tu}{Q^2}\,\log\frac{\td}{Q^2}~,\\
&&\nn\\
\DVtt &=& -2 -\frac{5\,\mg\,\tu\,\sdt}{2\,\mt^3} + 6 \,\log^2\frac{\t}{Q^2}\nn\\
&& +\frac{4\,\g}{\Delta}\,\left[
\g-\t-\tu + \frac{\mt\,\sdt}{\mg}\,\left(\t-\g-\tu\right)\right)
\,\log\frac{\t}{Q^2}\nn\\
&& -\frac{4\,\g}{\Delta}\,\left[
\g-\t+\tu + \frac{\mg\,\sdt}{\mt}\,\left(\t-\g+\tu\right)\right)
\,\log\frac{\g}{Q^2}\nn\\
&&+\left[\frac{8\,\g\,\tu}{\Delta} + \frac{2\,\mg\,\tu\,\sdt}{\mt^3}
\,\left(1-\frac{2\,\t}{\Delta}\,(\t+\g-\tu)\right)\right]
\,\log\frac{\tu}{Q^2}\nn\\
&&-\left(2-\frac{\mg\,\tu\,\sdt}{2\,\mt^3}\right)
\,\log\frac{\t}{\tu}\,\log\frac{\g}{Q^2}
-\left(2+\frac{\mg\,\tu\,\sdt}{2\,\mt^3}\right)
\,\log\frac{\t}{Q^2}\,\log\frac{\tu}{Q^2}\nn\\
&&+\frac{\mg\,\sdt}{2\,\mt^3}\,\left(\g+3\,\t\right)\,
\log\frac{\t}{\g}\,\log\frac{\tu}{Q^2}\nn\\
&&-\frac{2}{\Delta\,\t}\biggr\{
\frac{\mg\,\sdt}{\mt^3}\,\left[\frac{\Delta^2}{4}
+\t\,\left(\g-2\,\t+\tu\right)\,\Delta +\tq\,\left(\g-\t+\tu\right)^2
\right]\nn\\
&&~~~~~~
-\tu\,\left(\Delta+(\g+\t)\,(2\,\t-\tu)\right)-\left(\g-\t\right)^3\biggr\}
\,\Phi(\tu,\g,\t)\nn\\
&+& \biggr\{\tul \rightarrow \tdl\,,~~~ \sdt \rightarrow -\sdt \biggr\}~,\\
&&\nn\\
\label{DVcdtqt}
\DVcdtqt &=&-\frac{\mg}{2\,\mt\,\sdt}\,\biggr\{
5\,\tu
-4\,\tu\,\log\frac{\tu}{Q^2} +\left(\g-3\,\t\right)
\,\log\frac{\g}{\t}\,\log\frac{\tu}{Q^2}\nn\\
&&~~~~~~~~~~~~~~~+\tu\left(\log\frac{\tu}{\g}\,\log\frac{\t}{Q^2}
+\log\frac{\g}{Q^2}\,\log\frac{\tu}{Q^2}\right)\nn\\
&&~~~~~~~~~~~~~~~+\left[\frac{\Delta}{\t}-2\left(\g-\t+\tu\right)\right]
\Phi(\tu,\g,\t)\biggr\}\nn\\
&+& \biggr\{\tul \rightarrow \tdl\,,~~~ \sdt \rightarrow -\sdt \biggr\}~,
\eea
where $Q$ is the renormalization scale at which the $\drbar$
parameters entering the one-loop part of the corrections are
expressed, the function $\Phi(x,y,z)$ is defined in appendix D, and we
have used the shortcut $\Delta \equiv \Delta(\tu,\g,\t)$, where the
function $\Delta(x,y,z)$ is also defined in appendix D. We recall that
the derivatives of $\Delta V$ are computed at the minimum of the
effective potential, therefore the r.h.s.~of
eqs.~(\ref{DVcdtq})--(\ref{DVcdtqt}) is expressed in terms of
field-independent parameters (including the mixing angle $\theta_t$,
with $-\pi/2 <\theta_t<\pi/2$). Finally, the derivatives of $\Delta V$
that involve $\td$ can be obtained from
eqs.~(\ref{dvdtu})--(\ref{dvdtucdt}) by means of the replacements
$\tul\leftrightarrow\tdl$ and $\sdt\rightarrow-\sdt$.

\end{appendletterC}


\section*{Appendix D: two-loop functions}
\begin{appendletterD}
We provide in this appendix the explicit formulae for the two-loop
functions appearing in the $\oatas$ corrections to the Higgs mass
matrices. Differently from the approach of ref.~\cite{dsz}, we choose
to renormalize the two-loop effective potential {\em before} taking
its derivatives. As first shown in ref.~\cite{jackjones}, this is
equivalent to using the ``minimally subtracted'' two-loop functions:
\bea
\label{expJ}
J(x,y) & = & x\,y
\,\left(1-\lnb x\right)\left(1-\lnb y\right)~,\\
&&\nn\\
I(x,y,z) & = &
\frac12\,\left[(x-y-z)\,\lnb y\,\lnb z
+ (y-x-z)\,\lnb x\,\lnb z\nn + (z-x-y)\,\lnb x\,\lnb y \right]\\
&-&\frac52\,\left(x+y+z\right) +2\,\left(x\,\lnb x + y\,\lnb y
+ z\,\lnb z\right) - \frac{\Delta(x,y,z)}{2\,z} \,\Phi(x,y,z)~,\\
&&\nn\\
L(x,y,z) & = & J(y,z)-J(x,y)-J(x,z)-(x-y-z)\,
I(x,y,z)\,.
\eea

\noindent
In the above formulae, $\lnb x$ stands for $\log(x/Q^2)$, where
$Q$ is the renormalization scale. The functions $\Delta$ and $\Phi$
read, respectively,
\bea
\label{defdelta}
\Delta(x,y,z) & = & x^2 + y^2 + z^2 - 2\,(x y + x z + y z)\, ,\\
&&\nn\\
\label{defphi}
\Phi\,(x,y,z) & = & \frac{1}{\lambda} \left[
2\,\log x_+\,\log x_- - \log u \,\log v -
2\, \biggr( {\rm Li}_2 (x_+) + {\rm Li}_2 (x_-) \biggr) + 
\frac{\pi^2}{3} \right] \, ,
\eea
where ${\rm Li}_2(z) = -\int_0^z {\rm d}t \left[\log(1-t)/t\right]$ 
is the dilogarithm function and the auxiliary (complex) variables are:
\be
u = \frac{x}{z}\,,\;\;\;\;\;\;
v = \frac{y}{z}\,,\;\;\;\;\;\;
\lambda = \sqrt{(1-u-v)^2 - 4 \,u\, v}\,,\;\;\;\;\;\;
x_{\pm} = \frac{1}{2}\,\left[1 \pm (u-v) - \lambda\right] \, .
\ee
The definition (\ref{defphi}) is valid for the case $x/z < 1$ and 
$y/z < 1$. The other branches of $\Phi$ can be obtained using the symmetry 
properties:
\be
\label{symm}
\Phi\,(x,y,z) = \Phi\,(y,x,z)\,,\hspace{1cm}
x\,\Phi\,(x,y,z) = z \, \Phi\,(z,y,x) \, .
\ee
Finally, the following recursive relation for the derivatives of
$\Phi$ proves very useful for obtaining compact analytical results:
\be
\label{dphidx}
\Delta(x,y,z)\,\frac{\partial \,\Phi(x,y,z)}{\partial\, x} 
= (y+z-x)\,\Phi(x,y,z) + 
\frac{z}{x}\,\left[ (y-z)\,\ln\frac{z}{y}
+ x\,\left( \ln\frac{x}{y} + \ln\frac{x}{z}\right)\,\right]\, .
\ee
The derivatives of $\Phi$ with respect to $y$ and $z$ can be obtained
from the above equation with the help of the symmetry properties of 
Eq.~(\ref{symm}).

\end{appendletterD}


\end{document}